\documentclass[useAMS,usenatbib]{mn2e}
\usepackage{graphicx}
\usepackage{rotating,times,pictex,graphicx,latexsym}
\usepackage{color}
\usepackage{longtable,amsmath}
\usepackage{lscape}
\usepackage{lipsum}

\title[YSO jets in the Galactic Plane]{YSO jets in the Galactic Plane from
UWISH2:\\ II - Outflow Luminosity and Length distributions in Serpens and
Aquila} 

\author[Ioannidis \& Froebrich]{G.~Ioannidis$^{1}$\thanks{E-mail:
gi8@kent.ac.uk}, D.~Froebrich$^{1}$\thanks{E-mail: df@star.kent.ac.uk}\\ $^1$
Centre for Astrophysics and Planetary Science, University of Kent, Canterbury,
CT2 7NH, UK } 

\begin{document}

\date{Received sooner; accepted later}
\pagerange{\pageref{firstpage}--\pageref{lastpage}} \pubyear{2011}
\maketitle

\label{firstpage}

\begin{abstract}

Jets and outflows accompany the mass accretion process in protostars and young
stellar objects. Using a large and unbiased sample, they can be used to study
statistically the local feedback they provide and the typical mass accretion
history. Here we analyse such a sample of Molecular Hydrogen emission line
Objects in the Serpens and Aquila part of the Galactic Plane. Distances are
measured by foreground star counts with an accuracy of 25\,\%. The resulting
spacial distribution and outflow luminosities indicate that our objects sample
the formation of intermediate mass objects. The outflows are unable to provide a
sizeable fraction of energy and momentum to support, even locally, the
turbulence levels in their surrounding molecular clouds. The fraction of parsec
scale flows is one quarter and the typical dynamical jet age of the order of
10$^4$\,yrs. Groups of emission knots are ejected every 10$^3$\,yrs. This might
indicate that low level accretion rate fluctuations and not FU-Ori type events
are responsible for the episodic ejection of material. Better observational
estimates of the FU-Ori duty cycle are needed.

\end{abstract}


\section{Introduction}

Star formation and in particular the mass accretion process is accompanied by
the ejection of jets and outflows. These interact with the surrounding
interstellar medium by shocks which excite, ionise or dissociate atoms and
molecules. It is thought that these outflows provide localised feedback, i.e.
they infuse energy and momentum into the ISM. In particular in low mass star
forming regions where massive stars and their energetic radiation and winds are
absent, they might be the governing mechanism to terminate further star
formation (\cite{Walawender2005}). Thus, there are a number of 'small' scale
studies to characterise the population of outflows in nearby individual low mass
star forming regions (e.g. \cite{Stanke2001}, \cite{Walawender2005},
\cite{Hatchell2007}, \cite{Davis2009}, \cite{Khanzadyan2012}). 

However, a large fraction, if not the majority of Galactic star formation is
occurring in the presence of more massive stars and potentially in clusters
along the Galactic Plane. We hence aim to characterise the general population of
outflows from protostars and young stellar objects in an unbiased way. In order
to have a representative sample of outflows which is free from selection effects
we are conducting an unbiased search for jets and outflows in the Galactic Plane
using the UKIRT Wide Field Infrared Survey for H$_2$ (UWISH2,
\cite{Froebrich2011}). The sample of outflows from this survey will allow us to
perform a statistical investigation of their properties and to address some of
the open questions in the field such as: Why does a fraction of objects have no
outflows, i.e. what triggers/stops outflow activity? Are the outflows related to
FU-Ori type outbursts and if so how?

In this project, we focus our attention on the Serpens/Aquila region in Galactic
plane, covered by UWISH2. In particular we investigate the area 18$^\circ$\,$< l
<$\,30$^\circ$, -1.5$^\circ < b <$\,+1.5$^\circ$ which approximately covers 33
square degrees. In \cite{Ioannidis2012} (Paper\,I hereafter) we presented the
data (also discussed in \cite{Froebrich2011}) and discuss our detection method
of jets/outflows and their potential driving sources. We did increase the sample
of known Molecular Hydrogen emission-line Objects (MHOs) 15-fold and
investigated their basic properties such as fluxes, apparent projected lengths
and spatial distribution. We find that the flows tend to cluster in groups of a
few (3\,--\,5) objects on scales of 5\,pc, larger than typical young clusters.
The scale height of the outflows with respect to the Galactic Plane is about
30\,pc, similar to massive young stars.

In this paper we discuss in detail how we measure and calibrate the distance to
the outflows in our sample (Sect.\,\ref{dataandanalysis}). In
Sect.\,\ref{results} we present our analysis, results and discussion. We
re-evaluate the distribution of objects with respect to the Galactic Plane
taking into account the measured distances. We then determine the statistically
corrected luminosity functions and the associated star formation rate. We
continue by presenting our investigation of the total jet energy and momentum
input from our outflows into the interstellar medium. Furthermore, we analyse
the outflow length distribution and the frequency of mass ejections.

In our forthcoming paper (Ioannidis \& Froebrich, in preparation, hereafter
Paper\,III), we will investigate in detail the driving sources and how their
properties (e.g. luminosity, age, accretion rates) relate to the outflow
parameters (e.g. luminosity, length).

\section{Data analysis}\label{dataandanalysis}

\subsection{Distance Determination}

The determination of physical properties of the outflows, such as luminosity or
length, requires us to know the distances to all objects in our sample. As has
been shown in Paper\,I, the vast majority (above 90\,\%) of outflows in our
sample are new discoveries. Thus, it is highly unlikely that we find objects
with a known distance associated to all our objects. Even if we find literature
distances to most of our objects, then they will most likely be measured by a
mix of different methods -- introducing biases. Finally, it is impractical to
measure radial velocities for all objects (similar to the approach used in the
Red MSX Source (RMS, MSX - Midcourse Space Experiment) survey by
\cite{Urquhart2008} to determine distances. Only eight of our objects do
actually coincide with RMS sources of known distance (they are indicated by a
$+$ sign in the main result table in Appendix\,\ref{appendix1}).  We thus
require a way to determine the distances to all our objects in a homogeneous and
unbiased way in order to obtain e.g. a statistically correct luminosity
distribution. 

\begin{figure}\centering
\includegraphics[width=8.0cm]{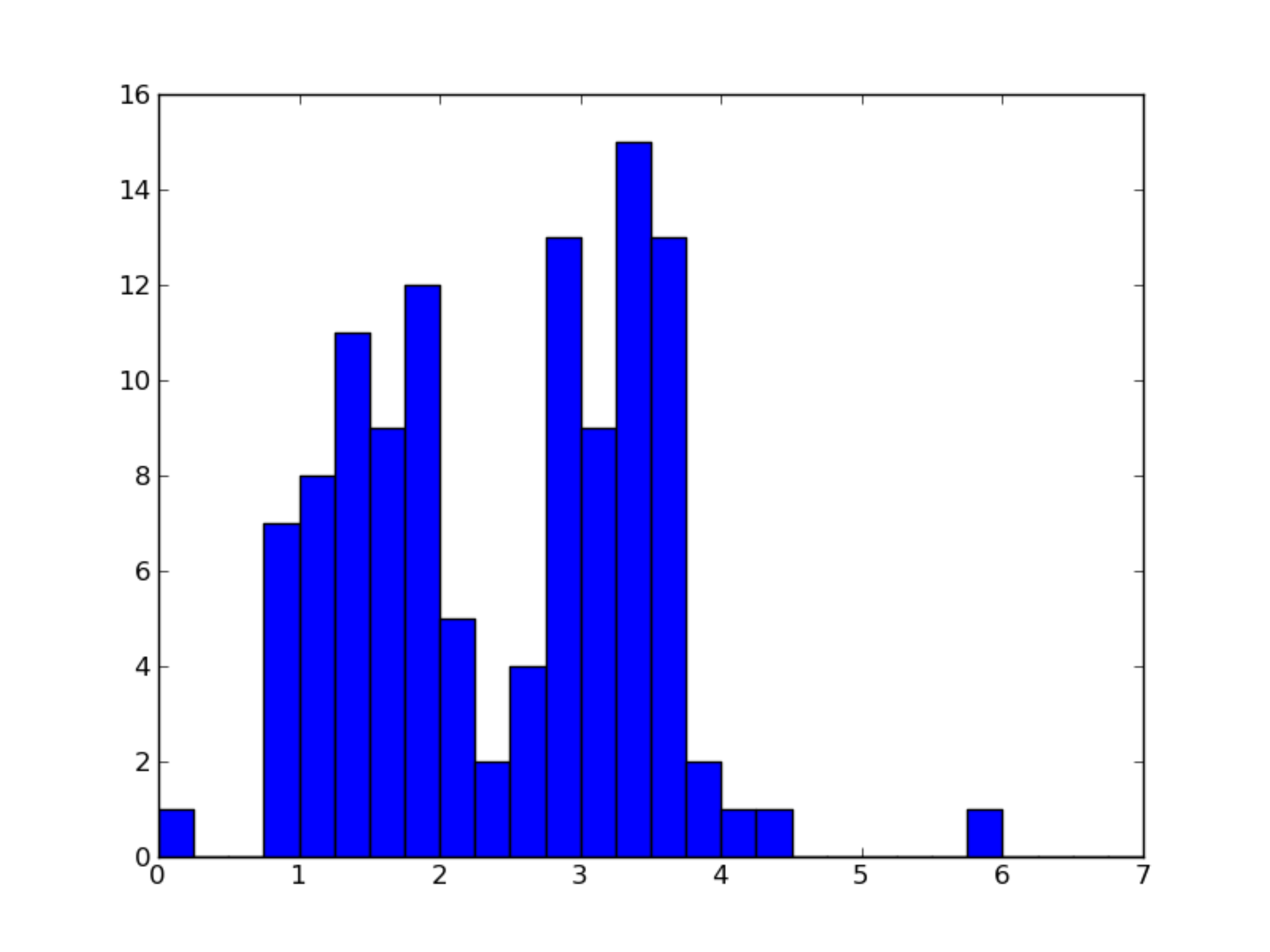}
\caption{\label{jklim} Distribution of the $J-K$ UKIDSS colour of stars near the
Glimpse source G024.1838$+$00.1198, one of our calibration regions. One can
clearly identify the separation of foreground and background stars at a value of
$J-K = 2.3$\,mag.}
\end{figure}

We use the UKIDSS GPS near infrared JHK data (\cite{Lucas2008}) to determine the
projected number density of foreground stars to the dark clouds associated with
our jets and outflows. A similar approach has been used recently in
\cite{Foster2012}. They find that this extinction distance method agrees well
with the maser parallax distances (within the errors) and is thus highly
reliable. Similarly \cite{mercer14} have used this method successfully to
determine the distance to the cluster Mercer\,14. 

For each cloud with detected jets and outflows we perform the following: i) We
select JHK photometry of the 'darkest' part of the cloud to ensure that as few
background stars as possible are included. The area selected has to be as large
as possible to get a good reliable estimate of the number of foreground stars.
ii) We plot a histogram of the $J-K$ colour of all selected stars and manually
identify the break ($J-K$)$_{lim}$ in the distribution caused by the cloud's
extinction. In Fig.\,\ref{jklim} we show an example of the cloud near
G024.1838+00.1198 (one of our calibration objects, see Sect.\,\ref{sect_cal})
for which we find ($J-K$)$_{lim} = 2.3$\,mag. We consider the break to be real
if at least 5\,--\,10 stars are apparently missing in one or more histogram
bins. iii) We determine the projected foreground star density as the number of
stars bluer than ($J-K$)$_{lim}$ per unit area. Note that only stars above the
local GPS completeness limit (determined as peak in the luminosity function for
the stars selected within the cloud) in every filter are included. iv) We use
the Besancon Galaxy model (\cite{Robin2003}) to determine out to which distance
we should expect the same number of foreground stars per unit area with the
local photometric limits. v) We estimate the distance uncertainty based on the
uncertainty of the number $N$ of foreground stars used in each field ($\Delta N
= \sqrt{N}$).

There are a number of MHOs (about 10\,\%) for which the described technique does
not work successfully. In these cases the objects are not seen in projection
onto an obvious dark foreground cloud. These outflows are either part of a very
distant cloud or have formed in a low $A_V$ region. For all these objects the
mean distance of all other outflows has been adopted in order to determine their
luminosity and lengths. In Table\,\ref{mainresults} in Appendix\,\ref{appendix1}
these MHOs are marked by a \dag. Note that they are not used in any of the
statistical analysis of the luminosity and length distributions. 

\subsection{Distance Calibration}\label{sect_cal}

Our adopted distance determination method is a star counting technique in
combination with a Galactic model. It has been used successfully by a number of
authors, e.g. in \cite{Foster2012}, \cite{Knude2010}, \cite{mercer14}. However,
it is unclear if there are any systematic biases in applying this method. The
Besancon model, for example, assumes a standard interstellar extinction law of
0.7\,mag of $A_V$ per kpc distance (in agreement with the reddening towards old
stellar clusters in the galactic disk; \cite{Froebrich2010}) which could be
systematically different along our sight line leading to systematic shifts in
the measured distances. Furthermore, the method will determine the distance to
the first dark cloud along the line of sight. A fraction of our outflows could
actually be situated at a larger distance, if a certain percentage of objects is
situated along sight lines with overlapping clouds. We thus require to calibrate
our distance calculation method with a sample of objects with known distances
which are situated in dark clouds (similar to our sample), in order to establish
the reliability and accuracy of the method. This calibration will not just be
used to verify the method, but rather to estimate by which factor our distances
are wrong for which fraction of objects. The RMS source list from
\cite{Urquhart2008} has been selected for this purpose, as it represents the
best available sample of objects associated with dark clouds, statistically
distributed in a similar way to our outflows.

We selected all RMS sources within our survey area which have an estimated
distance or (if there is a near/far ambiguity from the radial velocities) a
near-distance between 2\,kpc and 6.2\,kpc. We then determine for each RMS source
the distance in exactly the same way as for our outflows. There where a number
of sources for which we could not measure a distance, since they did not
coincide with an obvious dark cloud. These are most likely distant objects and
we hence remove them from our sample. Note that they make up 10\,\% of the RMS
objects and that we could not measure a distance for the same fraction of our
outflows. In total we were able to determine the distance to 41 RMS sources in
our survey area. 

We perform a linear regression between our and the RMS distance which leads to
the following calibration relation:

\begin{equation}\label{eq1}
d^{cal,1}_{RMS}[kpc] = 0.59 \times d_{sc}[kpc] + 1.97 [kpc]
\end{equation}

We denote with $d^{cal,1}_{RMS}$ calibrated distance of the RMS objects and with
$d_{sc}$ the distance determined from foreground star counting and the Besancon
Model. Note that we will refer to this as distance calibration method\,1
throughout the paper. The root mean square ($rms$) standard deviation of this
calibration is 1.0\,kpc. We note that the scatter in the determined distances
are not introduced by our determination of the ($J-K$)$_{lim}$ values and the
associated foreground star density. Expressed in units of the uncertainties of
our distance errors, the deviations amount to eight standard deviations. Thus,
the scatter is caused either by uncertainties in the RMS distances, large scale
foreground clouds or low extinction (see discussion in
Sect.\,\ref{sect_statcorr}). Note that \cite{Urquhart2008} quote an error of
1\,kpc for their distances, based on peculiar motions of up to 10\,km\,s$^{-1}$.
This conservative estimate indicates that a large fraction of the scatter could
be intrinsic to the RMS distances. Thus, our distances might be more accurate
than implied by the $rms$-scatter, in agreement with the high accuracy found in
\cite{Foster2012}.

We construct a histogram of the logarithmic distance ratio $R = \log(d_{RMS} /
d^{cal,1}_{RMS})$ with $d_{RMS}$ the distance in the RMS catalogue. This
histogram has a width indicating a typical scatter of about 25\,\% for the
distances. 

Furthermore, we try to identify correlations of $d_{sc}$ with other outflow
related parameters such as the galactic coordinates or the ($J-K$)$_{lim}$ value
used in the distance calculation. Only the Galactic Longitude $l$ shows a
marginal correlation (correlation coefficient $r = 0.41$). All other parameters
have no systematic influence on the distance. If we hence consider the Galactic
Longitude $l$ in the calibration we find the following calibration relation:

\begin{equation}\label{eq2}
d^{cal,2}_{RMS}[kpc] = 0.69 \times d_{sc}[kpc] + 0.16 [kpc/deg] \times l [deg]
- 2.41 [kpc]
\end{equation}

Note that we refer to this as distance calibration method\,2 throughout the
paper.  The $rms$-scatter of the distances using this calibration is 0.9\,pc
(six times the uncertainties of the individual measurements), slightly smaller
than for calibration method\,1. Including $l$ in the calibration leads to, on
average, slightly higher distances and thus luminosities (and lengths) for our
outflows, but the effect is marginal.

 %
 %

To summarise, we assume that the distances to the RMS calibration sources are
accurate. Equations\,\ref{eq1} and \ref{eq2} are used to calibrate, in two ways,
the distances estimated to each outflow using foreground star counts. These
distances are listed  in Table\,\ref{mainresults} in Appendix\,\ref{appendix1}.

%
 %

\subsection{Statistical Corrections}\label{sect_statcorr}

The above discussed distance calculation and calibration method shows to what
extend our method over/underestimates distance to the RMS objects in our field.
The distribution the logarithmic distance ratio $R$ essentially represents the
probability distribution of the uncertainties in our distance determination (one
for each calibration method). If $R$ is positive we underestimate the distance
(e.g. due to foreground extinction/clouds), if $R$ is negative our distances are
too large (e.g. caused by low $A_V$ clouds and hence the inclusion of blue
background stars in the foreground star density). We now take the distribution
of uncertainties for the measured distances to the outflows in the survey field
is the same as for the calibration objects. This is justified, since most of the
RMS objects are young stellar objects and are distributed in the same distance
range and with the same scale height as our outflows (see
Sect.\,\ref{sect_distdistr} and \ref{sect_scaleheight}).

Thus, in order to determine e.g. the luminosity distribution for the outflows,
we add each outflow $N = 10000$ times into the luminosity distribution. Each
time the distance used in the calculation is the calibrated distance plus an
uncertainty that is drawn from a sample of random numbers with the same
distribution as $R$. Thus, we can determine statistically correct luminosity and
length distributions for our outflows. 

The luminosities are further influenced by interstellar and cloud extinction
local to the outflow. Hence, we have to apply a further correction for each
determined luminosity. Based on the distance for each object we use  the
standard 0.7\,mag of optical extinction ($A_V^d$) per kpc distance from
\cite{Froebrich2010}. We further estimate the local or cloud extinction $A_V^c$
at the position of the outflow using the extinction maps by \cite{Rowles2009}.
These maps give the total $A_V^{tot}$ along the line of sight and we take:

\begin{equation}
A_V^{tot} = A_V^{d} + A_V^{c}
\end{equation}

We do of course not know where in the cloud the molecular hydrogen emission is
situated (front or back). Note that the extinction values are low enough as to
not introduce any bias towards outflows situated at the front of the clouds.
Thus, for each of the $N$ times we add every outflow into the luminosity
distribution (see above) we use an extinction of $A_V^d + w \times A_V^c$ where
$w$ is a random number drawn from a homogeneous distribution between zero and
one. All $A_V$ values are converted into K-band extinction (i.e. the extinction
in the 1-0\,S(1) line) using $A_V = 9.3 \times A_K$ following \cite{Mathis1990}.

In order to establish a statistically correct length distribution, we need to
correct for the unknown inclination angle of the outflow. We find, however, that
it is more convenient not to apply this correction, but rather try to simulate
the projected length distribution since outflows almost perpendicular to the
plane of the sky are missing in our sample (see later in
Sect.\,\ref{sect_lengthdistr}).

\section{Results and Discussion}\label{results}

\begin{figure}\centering
\includegraphics[width=8.0cm]{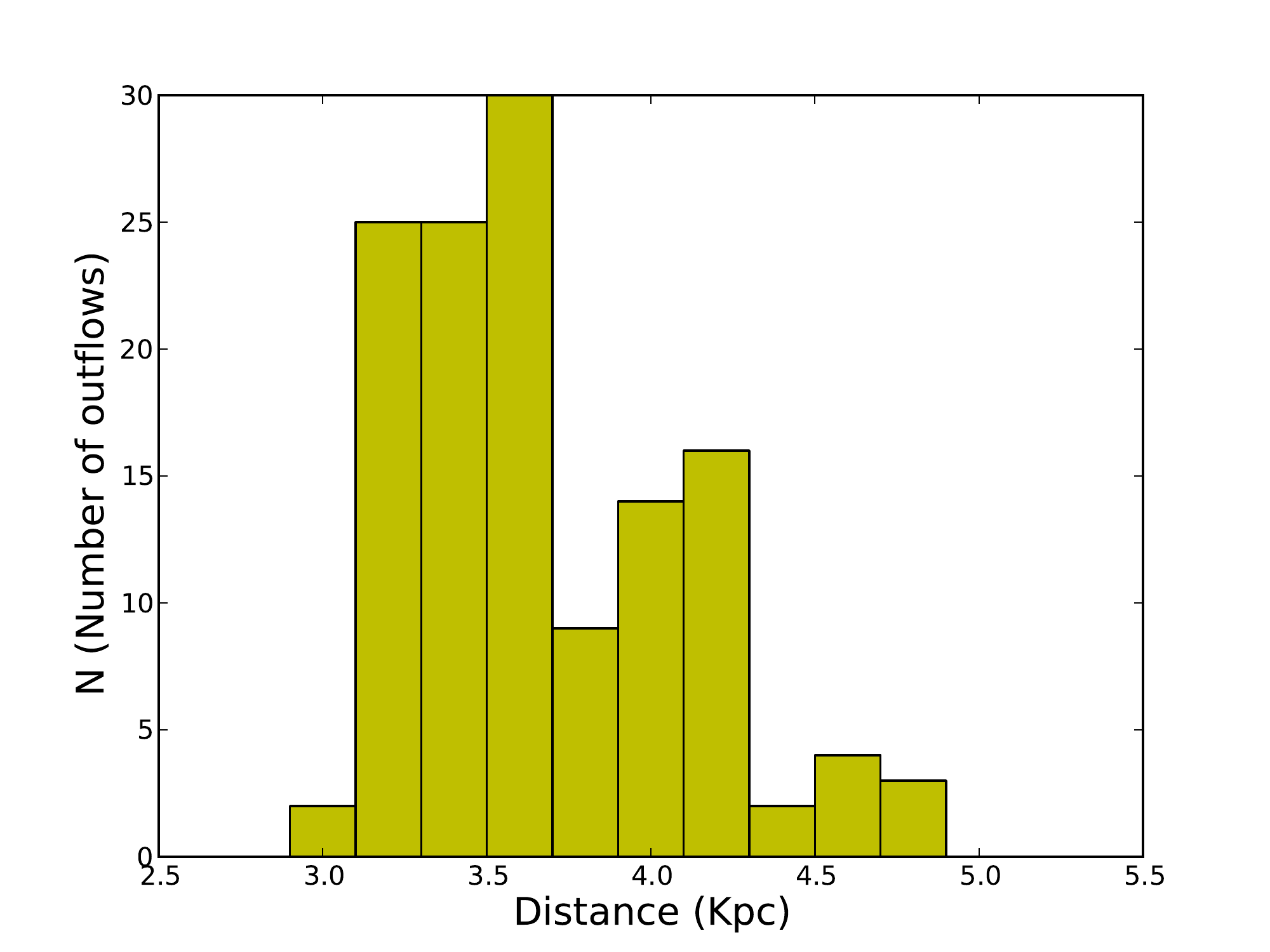}
\caption{\label{distdistr} Distribution of distances of our outflows (using
calibration method\,1). There is a clear peak between 3.0\,kpc
and 3.5\,kpc and a smaller, less obvious peak at about 4.2\,kpc. Objects without
determined distance are excluded.}
\end{figure}

\begin{figure}\centering
\includegraphics[width=8.0cm]{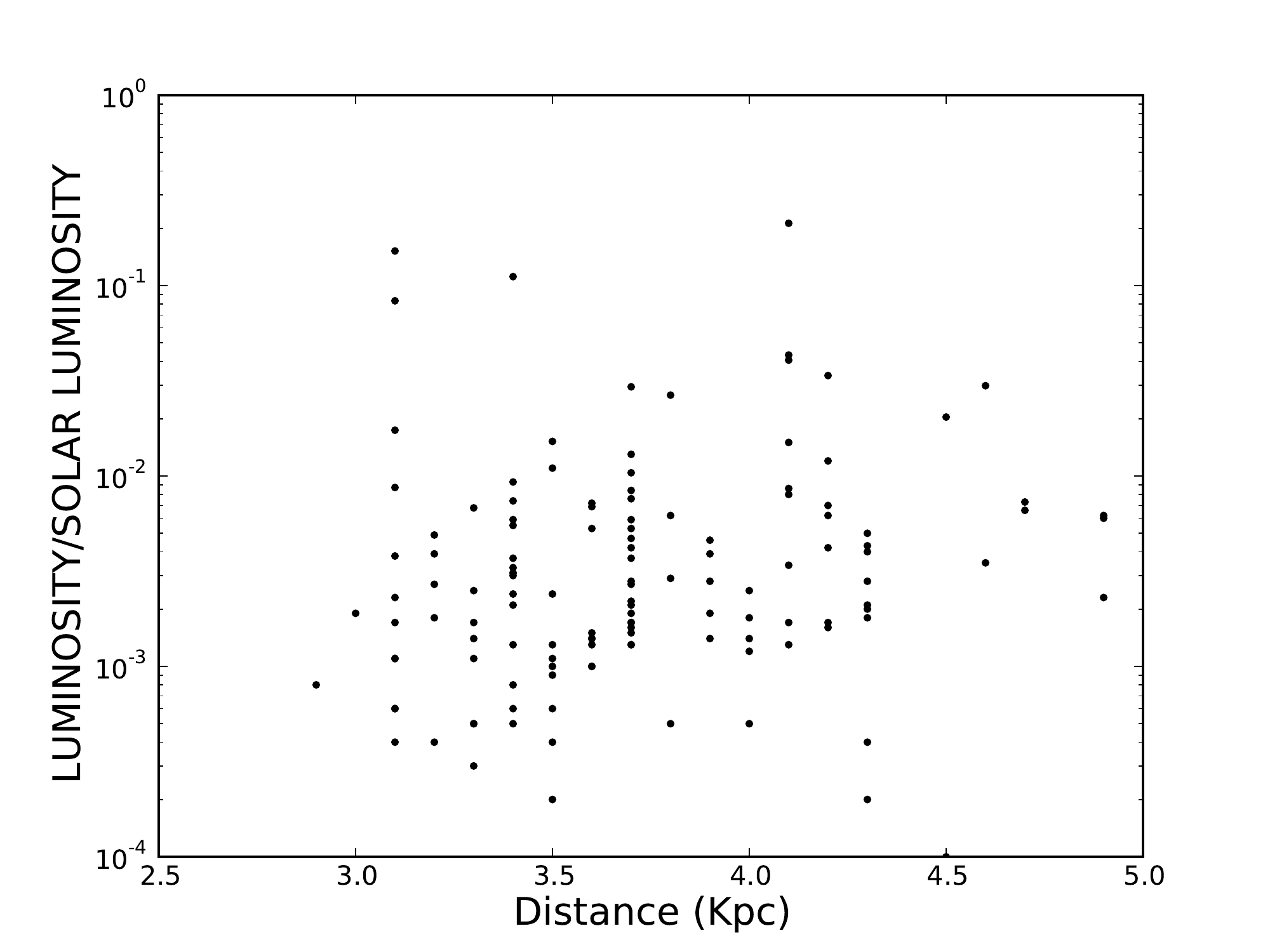}
\caption{\label{lumdist} Outflow distances vs. H$_2$ 1-0\,S(1) luminosities
(based on calibration method\,1). We detect outflows out to 5\,kpc, and there is no clear trend in the diagram. Our estimated completeness
limit for the outflow H$_2$ 1-0\,S(1) luminosity within 5\,kpc is
10$^{-3}$\,L$_\odot$. The large number of objects at 3.7\,kpc is due to the
objects where we could not determine any distance and applied the mean
distance.}   
\end{figure}
 
\subsection{Distance distribution}\label{sect_distdistr}

Since there are two calibration methods for the determined distance, we list
both values in Table\,\ref{mainresults} in Appendix\,\ref{appendix1}. In general
method\,1 (simple calibration without considering the Galactic Longitude) gives
slightly lower distances than method\,2. A $\dag$ symbol indicates outflows
where we could not determine a distance and we used the mean distance measured
for all other objects in those cases. Note that we exclude all these sources
from any further statistical analysis. 

The distances measured for our outflows are generally in the range from 2.0\,kpc
to 5.0\,kpc. This is within the distance range of the RMS objects that are used
for the calibration. The distribution of distances shows a peak at
3.0\,--\,3.5\,kpc, indicating the presence of a spiral arm along this sight
line. There is a second, less obvious peak at 4.2\,kpc (see
Fig.\,\ref{distdistr}). In \cite{Urquhart2011} a similar increase in the number
density of RMS sources in the same area can be seen at distances of about
3\,--\,4\,kpc. The difference in the number of objects in both peaks is not due
to completeness issues. The fraction of low luminosity outflows is the same for
both (see Fig.\,\ref{lumdist}). Thus, the more nearby feature has a larger
number of active outflow driving sources.

\begin{figure*}\centering
\includegraphics[width=17.5cm]{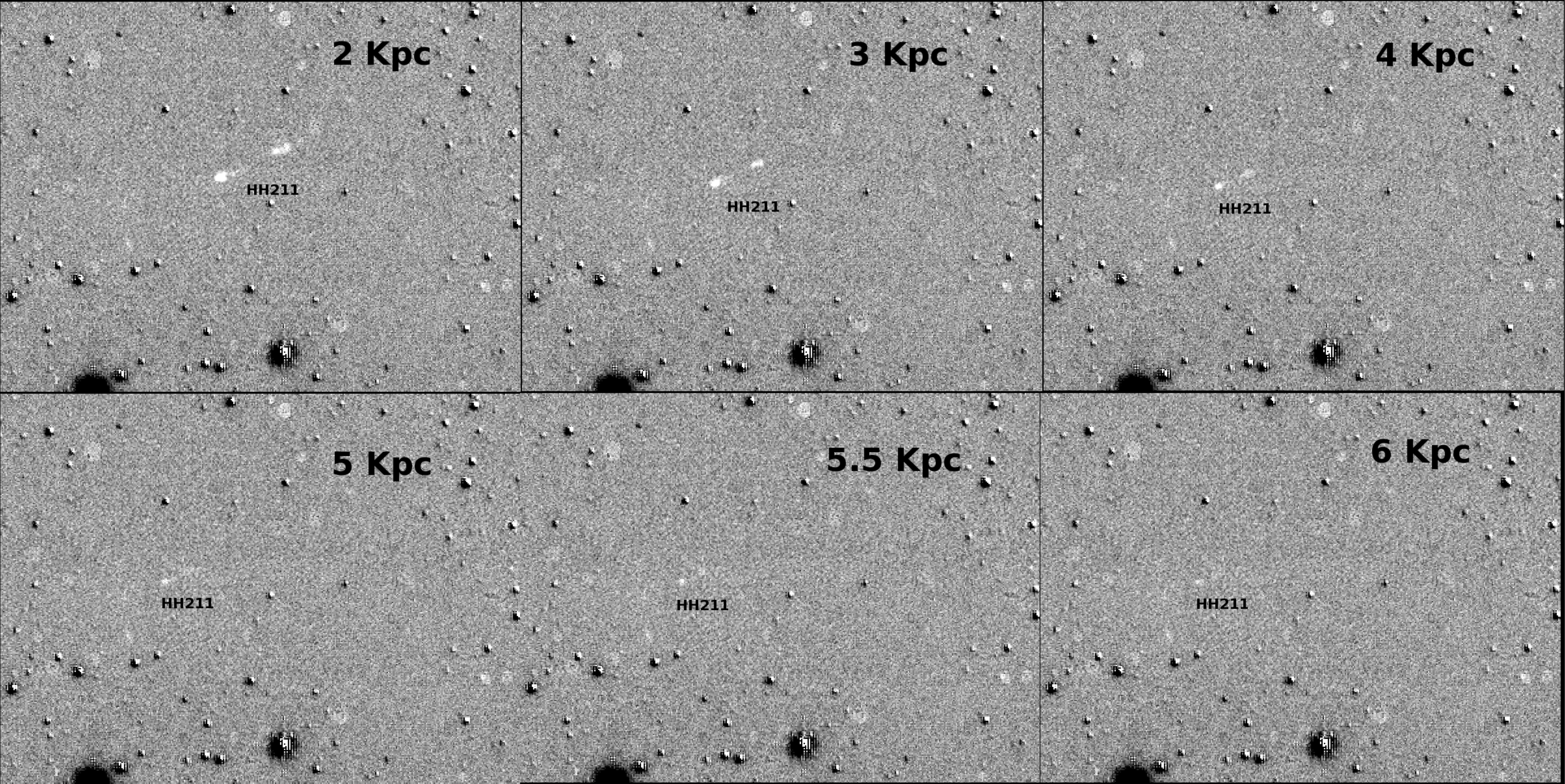}
\caption{\label{hh211} HH\,211 as it would appear in one of our H$_2$\,-\,K
difference images at different distances (near the 'HH\,211' label). The well
known object is placed flux calibrated and scaled in a relatively 'normal' cloud
region in one of our survey images. The small images are
80\arcsec\,$\times$\,60\arcsec\ in size and the distance of HH\,211 are
indicated in each panel. At distances at and above 5\,kpc the outflow
becomes indistinguishable from a very red or variable point source.} 
\end{figure*}

In order to estimate out to which distance we would in principle be able to
detect molecular hydrogen outflows we used HH\,211 as an example. This outflow
is driven by a young Class\,0 source and emits about
3.1\,$\times$\,10$^{-3}$\,L$_\odot$ in the 1-0\,S(1) line of H$_2$
(\cite{Eisloeffel2003}). We obtained a flux scaled image taken of this object in
the 1-0\,S(1) line from \cite{Eisloeffel2003} and placed it scaled to different
distances into one of our UWISH2 images (which represents a typical region).
Note that we do not apply any extinction corrections to the flux. The result of
the exercise is shown in Fig.\,\ref{hh211}. At distances below 5\,kpc the
outflow can easily be identified as bright extended emission with a clear
bipolar structure. Several of our outflows (e.g. MHO\,2256, 2289, 2292, 2441)
have a similar appearance. At larger distances, however, the apparent size of
the H$_2$ emission knots sinks below our spatial resolution and thus the
brightness decreases significantly. Furthermore, the emission line knots will
appear as red or variable point sources (which might remain in the H$_2$\,-\,K
difference images - see Paper\,I) and would thus most likely no be contained in
our outflow sample, in particular if it was situated in a region higher 
foreground star density.

In Fig.\,\ref{lumdist} we plot the outflow distances vs. the luminosity in the
1-0\,S(1) line of H$_2$ as measured by us. We see that low luminosity objects
are sparse, but there is no real trend of the minimum detected luminosity with
distance. Thus, based on Fig.\,\ref{lumdist} we conclude that our survey is
complete to 5\,kpc for objects with more than 10$^{-3}$\,L$_\odot$ in the
1-0\,S(1) line of H$_2$. This corresponds to HH\,211 like outflows with about
1\,mag of extinction in the K-band. The completeness limit is mostly set by the
flux detection limit of 3\,$\times$\,10$^{-18}$\,W\,m$^{-2}$ (discussed in
Paper\,I), since most our objects are extended up to a distance of 5\,kpc.

\begin{figure}\centering
\includegraphics[width=8.0cm]{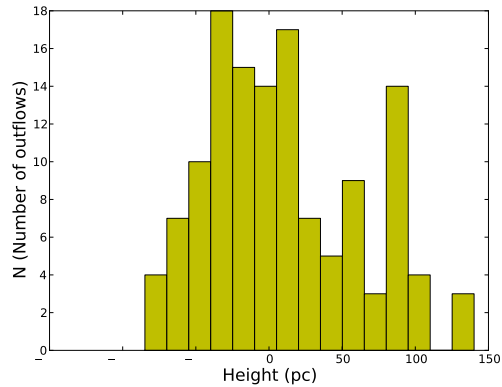}
\caption{\label{scaleheight} Distribution of the height above and below the
Galactic Plane of our outflows (based on calibration method\,1). The mean is
shifted to about 20\,pc below the Plane, and the width of the distribution shows
a scale height of approximately 30\,pc. Objects without distance determination
are excluded.}  
\end{figure}
 
\subsection{Outflow scale height}\label{sect_scaleheight}

With the distances for all outflows we are able to determine the distribution of
objects with respect to the Galactic Plane. We already investigated this
distribution in Paper\,I assuming a distance of 3\,kpc for all outflows. As
discussed above, most of our objects are indeed roughly at this distance, but
the average is about 3.5\,kpc. Thus, no significant change in the scale height
compared to Paper\,I is expected. Only a small increase of 15\,\% should occur.

Figure\,\ref{scaleheight} shows the height above and below the Galactic Plane
for all outflows based on distance calibration method\,1. The distribution is
shifted to about 20\,pc below the Galactic Plane indicating that in this region
of the survey most star forming clouds are at negative galactic latitudes. The
one sigma width of the distribution, or the scale height, is of the order of
30\,pc. Hence our objects show the same distribution as typical massive OB stars
(scale height about 30\,--\,50\,pc, \cite{Reed2000}, \cite{Elias2006}) and the RMS
sources in this area (\cite{Urquhart2011}). This justifies our use of the RMS
sources as distance calibrators, since their distances and height distributions
are the same as measured for our outflows. This in turn suggests that our
outflows and massive star formation in the Galactic Plane are linked, even if
only eight RMS objects coincide with any of our outflows (two of those might
actually be background sources). In the forthcoming Paper\,III we will show that
the driving sources for our outflows seem to be on average intermediate mass
sources.

\subsection{Driving source verification}

The purpose of this paper is to investigate the luminosity and length
distribution of the discovered outflows. Both rely on a correct as possible
identification of the driving sources. This 'subjective' task has been performed
as described in Paper\,I. In order to ensure the correctness of the source
identification we have repeated this task (half a year after it has been done
originally) for all MHOs discovered in our field. For the vast majority of
objects we have identified exactly the same objects as potential driving
sources. Only for five out of the 134\,MHOs did we select a different potential
source. 

We thus list in Appendix\,\ref{appendix2} and \ref{appendix3} the new properties
and finding charts of the MHOs with a different source candidate (same as main
tables in Paper\,I). In the light of these small changes we have re-done all the
analysis from Paper\,I and there are no changes to any of the results and
conclusions. We have also done all the analysis for this paper with both
datasets and again, there are absolutely no differences in any of the results
and conclusions. 

\subsection{Outflow luminosity function}\label{lumfunc}

\begin{figure}
\includegraphics[width=8.0cm]{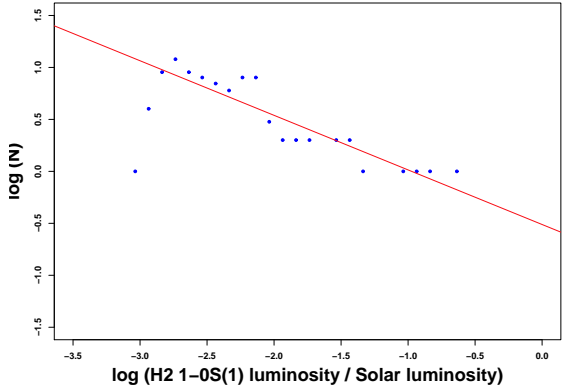}
\caption{\label{lumraw} 1-0\,S(1) Luminosity distributions of our outflows for
the distance calibration method\,1. All
objects with no measured distance are excluded, as are objects below the flux
completeness limit. The fitted power law slope (about -0.5 to -0.7) is
dependent on the histogram bin size due to the small number of objects.}
\end{figure}

The luminosities of our outflows (not corrected for extinction) in the 1-0\,S(1)
line of H$_2$ range from about 0.001 to 0.1\,L$_\odot$, and depend slightly on
the distance calibration method used. In Fig.\,\ref{lumraw} we show in a log-log
plot the luminosity function for distance calibration method\,1. The
corresponding plot for method\,2 and all other luminosity functions are
summarised in Appenix\,\ref{appendix4}. All objects below the flux completeness
limit and without properly determined distances are excluded. 

The luminosity distribution represents a power-law of the form:

\begin{equation}
N \propto L_{1-0 S(1)}^{\alpha}
\end{equation}
with $\alpha$ in the range from -1.5 to -1.7, depending on the distance
calibration method and histogram bin size.

\begin{figure}
\includegraphics[width=8.0cm]{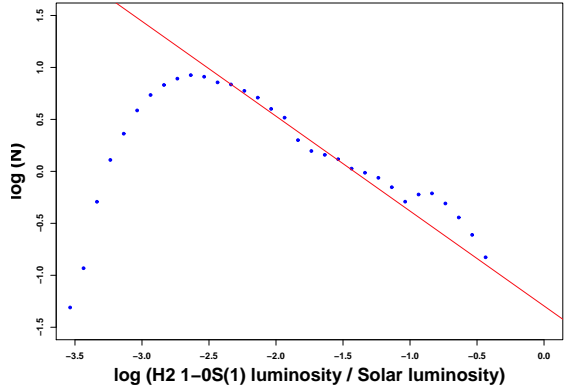}
\caption{\label{lumcor} 1-0\,S(1) Luminosity function of our outflows after we
corrected for the statistical uncertainties in the distance by calibration
method\,1. Each outflow has been placed $N=10000$ times into 
the histogram (see text for details). All objects with no measured distance are
excluded, as are objects below the flux completeness limit. The slopes for both
calibration methods are indistinguishable and have a value of -0.9. }
\end{figure}

When we apply our statistical distance correction to the luminosity
distribution, we obtain the luminosity function as shown in Fig.\,\ref{lumcor}.
The statistical consideration of our distance uncertainties does influence the
slope of the resulting luminosity function. It steepens the distribution, i.e.
statistically our sample contains more low luminosity objects. The luminosity
distributions for the distance calibration method\,1 and 2 are also power-laws
with slopes of $\alpha_1 = -1.89$ and $\alpha_2 = -1.88$, independent of the
histogram bin width. Essentially, the two distributions are indistinguishable
from each other. Thus, the shape of the luminosity distribution does not depend
on the detailed way we calibrate our distance. Only the absolute values for the
luminosities change slightly. 

We finally apply the extinction correction based on the distance and the local
cloud extinction. The resulting luminosity distributions are shown in
Appendix\,\ref{appendix4} and are almost identical. The power-law slopes
slightly increase to $\alpha_1 = -1.93$ and $\alpha_2 = -1.95$. 

In summary, the number distribution of the 1-0\,S(1) luminosities of our
outflows can be represented by a power-law with a slope of $\alpha = -1.9$ with
an uncertainty of about 0.1. This uncertainty accounts for possible changes of
the slopes due to the extinction correction.

Based on some simple, statistically correct assumptions, we can investigate how
such an outflow luminosity function can be interpreted. i) The total flux
emitted by the outflow is ten times larger than the 1-0\,S(1) flux. Hence the
measured outflow luminosities are proportional to the total outflow luminosity
emitted in all molecular hydrogen lines. This is correct for shocks at about
2000\,K (\cite{CarattioGaratti2006}), and has also been used by many other
authors in statistical calculations (e.g. \cite{Stanke2002}); ii) Most of our
outflows are driven by young protostars. The fact that we detect sources for
only half the MHOs supports this fact. Furthermore, many sources are only
detected at mid infrared wavelengths. We will present a detailed analysis of the
source properties in Paper\,III, where we show that the driving sources are
young embedded objects of intermediate mass.

We can use the empiric relationship of the outflow H$_2$ luminosity and the
bolometric driving source luminosity from \cite{CarattioGaratti2006}.

\begin{equation}
\log(L_{H_2}) = 0.58 \times \log(L_{bol}) - 1.4.
\end{equation}

Thus, we find for the distribution of the driving source bolometric
luminosities:

\begin{equation}
N \propto L_{H_2}^{-1.9 \pm 0.1} \propto L_{bol}^{-1.10 \pm 0.05}
\end{equation}

If all our sources are protostars, then $L_{bol}$ is dominated by the accretion
luminosity $L_{acc}$ which scales like: $L_{acc} \propto \dot{M} M R^{-1}$. We
can either use that the accreting central core has a constant density (then $R
\propto M^{1/3}$ and thus $M R^{-1} \propto M^{2/3}$) or a constant radius
(following \cite{Hosokawa2011} and thus $M R^{-1} \propto M$). For this range of
possibilities the accretion luminosity will thus scale as $L_{acc} \propto
\dot{M} M^{0.85 \pm 0.15}$. 

The mass accretion rate could scale as a power law with mass ($\dot{M} \propto
M^\beta$). We observe each object at a time when it has accreted a fraction $X$
of its final mass, which statistically should be the same for all objects.
Lastly, the distribution of final masses of the sources of our outflows should
represent a Salpeter like mass function. Hence we find that

\begin{equation}
N \propto M^{-2.35} \propto M^{-(1.1 \times \beta + 0.95) \pm 0.20}
\end{equation}

and thus $\beta = 1.3 \pm 0.2$. Which means that based on our outflow luminosity
function, the average mass accretion rate for protostars scales like

\begin{equation}
\dot{M} \propto M^{1.3 \pm 0.2}. 
\end{equation}

Finally, the accretion time scale $t_{acc}$ for an object of mass $M$ would
scale like

\begin{equation}
t_{acc} \propto M^{-0.3 \pm 0.2},
\end{equation}

i.e. more massive stars spend less time accreting material. Note that these
results have been determined based on a number of assumptions and hence might be
slightly different in reality. However, based on our data we can certainly rule
out that the average mass accretion rates for protostars driving our outflows is
independent of the final stellar mass. Any further details such as the exact
values and uncertainties of the inferred power law index should be investigated
with more detailed numerical models able to link accretion rates, source and
outflow luminosities (e.g. \cite{Smith1998}). 

If we would not correct our luminosity distribution for the distribution of
uncertainties ($R$), then we would obtain $\beta = 1.8$, an even stronger
dependence of the average mass accretion rate on the final stellar mass. 

We can also compare our result to the data from \cite{Stanke2002} who
investigated the outflow luminosities in Orion\,A. There the 1-0\,S(1)
luminosities span a range from 10$^{-4}$ to 10$^{-2}$\,L$_\odot$ and are hence
one order of magnitude smaller on average than our values. The distribution of
the L$_{H_2}$ values is flatter than ours, with a value of $\alpha = -1.1$,
which would lead (with the same assumptions as above) to $\beta = 2.8$. 

\subsection{Star Formation Rate}

When we convert our 1-0\,S(1) luminosities into an H$_2$ luminosity (without
accounting for extinction), our outflows cover a range of brightnesses from
0.01\,L$_\odot$ to 1.0\,L$_\odot$. This is in good agreement with the values for
other molecular hydrogen outflows e.g. in \cite{CarattioGaratti2008}. In this
paper one can also see that only very few objects are brighter than
1.0\,L$_\odot$ and thus most of our outflows are driven by low and/or
intermediate luminosity/mass protostars.

The total H$_2$ outflow luminosity of all objects in our investigated area is
9\,L$_\odot$ or 12\,L$_\odot$, depending on the distance calibration method. If
we correct for extinction using $A_K = 1$\,mag (a typical value for the objects
in our field), then the total H$_2$ luminosity in the survey area is
25\,L$_\odot$, or 10\,L$_\odot$/kpc$^2$.

With the assumptions from Sect.\,\ref{lumfunc} (the outflow luminosity is linked
to the accretion luminosity of the driving protostars as shown in
\cite{CarattioGaratti2008}) this converts into a total of
6\,$\times$\,10$^4$\,L$_\odot$ of accretion luminosity. Note that this will be a
lower limit, since there are some objects in \cite{CarattioGaratti2008} which
have much higher source luminosities compared to the general trend. If  a
typical protostar in our sample accretes onto a 2\,M$_\odot$ intermediate mass
core of 1.5\,R$_\odot$ (\cite{Hosokawa2011}) then we can determine the total
mass accretion rate in the survey area. If we normalise this to the area of 2.6
square kiloparsec we find a limit for the star formation rate ($SFR$) per square
kiloparsec in the galactic disk.

\begin{equation}
SFR > 2 \times 10^{-3} M_\odot yr^{-1} kpc^{-2}
\end{equation}

Our survey region covers an area roughly 4\,--\,7\,kpc from the Galactic Centre.
According to \cite{Boissier1999} the star formation rate in the Milky Way
($SFR_{MW}$) drops significantly at galactocentric distances above 8\,kpc. Thus,
if we scale up our value to 200\,square kiloparsec we find a limit of

\begin{equation}
SFR_{MW} > 0.4 M_\odot yr^{-1}
\end{equation}

This is in agreement with recent estimates for the Galactic star formation rate
e.g. by \cite{Robitaille2010} who found 0.7\,--\,1.5\,M$_\odot$\,yr$^{-1}$ based
on the analysis of Spitzer detected young stellar objects. In Paper\,III we will
estimate the properties of the driving sources in more detail, which will allow
us to determine a more accurate limit for $SFR_{MW}$.

 %
 %

\subsection{Outflow energetics}

We can also investigate the total jet energy and momentum input from our
outflows into the interstellar medium. As we have seen above, the typical object
in our sample is a jet from a low and/or intermediate mass star. Furthermore,
our data does not allow us to directly measure the jet power. We thus apply the
method and generic values used in \cite{Davis2008} in order to get an order of
magnitude estimate. Hence, we use $4 \times 10^{37}$\,J as the typical energy
input of each jet and 1\,M$_\odot$\,km\,s$^{-1}$ as momentum input. 

The turbulent energy in a cloud is approximately the cloud mass times the square
of the turbulent velocity dispersion. For the latter we take 1\,km\,s$^{-1}$ as
a typical value, since the energy input from jets and outflows occurs usually
locally (within at most a few parsec) from the star formation site, hence in
regions where the turbulent velocities are not extremely high. Note that this
value is also typical for nearby GMCs such as Perseus (\cite{Davis2008}). 

Thus, the total energy input from our 130 outflows can provide enough turbulent
energy for a total mass of just $2.5 \times 10^3$\,M$_\odot$. We can compare
this to the total molecular gas mass within our survey region. According to
\cite{Casoli1998} one expects about 3\,M$_\odot$\,pc$^{-2}$ of molecular gas
(\cite{Boissier1999} predict similar value of 6\,M$_\odot$\,pc$^{-2}$ at the
galactocentric radius of our objects), which adds up to a total of
10$^7$\,M$_\odot$ in our survey area. However, as noted earlier, the energy
input from the outflows will only occur locally, i.e. in the high density
regions. These are most likely the parts of the cloud where the (column) density
is above the star formation threshold. \cite{Rowles2009} found that typically
just one percent of a cloud is at these densities. Thus, only 10$^5$\,M$_\odot$
of cloud needs to be considered (this does not explain where the turbulent
energy in the low column density regions originates). 

In any case, the amount of mass that can be supported by the jets discovered in
the survey area is a factor of 40 smaller than the actual mass at high
densities. Thus, only if there are many generations of jets and outflows in each
star forming region would they provide enough energy input to account for the
turbulent energy. A typical age scatter of two million years would only allow
ten generations of protostars. Thus, even locally, in the high density star
forming fraction of the clouds, feedback from jets and outflows is insufficient
as a source of the turbulent energy. Hence, high star formation rate regions
like NGC\,1333, which seem able to locally inject enough momentum to support the
cloud (\cite{Walawender2005}) are not common in our survey area. 

\begin{figure}
\includegraphics[width=8.0cm]{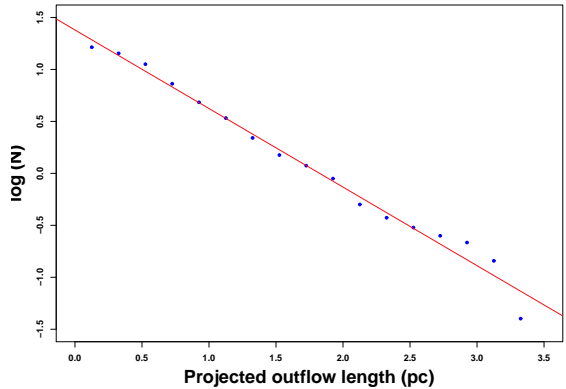}
\caption{\label{lengthdistribution} Statistically corrected distribution of the
projected lengths of our outflows for the distance calibration method\,1. Note
that the distribution resembles an exponential and not a power law.} 
\end{figure}

\subsection{Outflow length distribution}\label{sect_lengthdistr}

The projected lengths have been calculated for all outflows with an identified
driving source candidate. The lengths, derived using both distance calibration
methods are listed in Table\,\ref{mainresults} in Appendix\,\ref{appendix1}.
Note that we do not apply any corrections for single sided outflows, to allow a
comparison to other works (e.g. \cite{Stanke2002}, \cite{Davis2008},
\cite{Davis2009}). We find a steep decrease in the number of flows with
increasing length. In our sample we have between 15\,\% and 18\,\% of objects
with a projected length above 1\,pc (depending on the adopted distance
calibration). If we apply a statistical correction of $4/\pi$ for a random
distribution of inclination angles, then the fraction of parsec scale flows
increases to about 25\,\%. Compared to other surveys the fractions of parsec
scale flows (uncorrected for inclination) are: i) in Orion\,A \cite{Stanke2002}
8\,\%; ii) in Taurus, Auriga, Perseus \cite{Davis2008} 12\,\%; iii) in Orion\,A
\cite{Davis2009} 9\,\%. Thus, since our survey traces more luminous outflows
(see above), the fraction of parsec scale flows seems to be higher for brighter
objects.

In Fig.\,\ref{lengthdistribution} we show the distribution of the projected
jet/outflow lengths in our sample. The plot is corrected for our statistical
uncertainties in the distance calculations for method\,1. However, both
distributions are extremely similar and show an exponential decrease in the
number of objects with increasing length and not a power law behaviour. The
slope in the diagram hence indicates that the number $N$ of outflows is related
to the flow length in the following way:

\begin{equation}
N \propto 10^{- 0.75 \times lenght [pc]}
\end{equation}

We have run some simple simulations in order to understand the observed
projected lengths distribution. As already stated in Paper\,I, simply assuming
all jets are of the same length and randomly orientated should result in a
completely different distribution (more larger than shorter flows up to a
maximum projected length). A model of randomly orientated jets with uniformly
distributed  ages and a constant velocity fails to reproduce the data, in the
same way as using a constant age and uniformly distributed velocities.

We therefore developed a family of models based on jets with different ages,
homogeneously distributed between a minimum $a_{min}$ and maximum $a_{max}$.
Furthermore, the jet velocities also range from a minimum $v_{min}$ to maximum
$v_{max}$ value which are homogeneously distributed. Finally, the outflow
inclination angle (angle between the jet axis and the line of sight) ranges from
a minimum $i_{min}$ to 90$^\circ$. 

We then generated 16000 samples of 68 jets (the same number as jet lengths in
our data), with parameters selected randomly from the following ranges:

{\centering

$1000yrs \le a_{min} \le 5000yrs$

$10000yrs \le a_{max} \le 30000yrs$

$0 km/s \le v_{min} \le 50 km/s$

$90 km/s \le v_{max} \le 150 km/s$

$0^\circ \le i_{min} \le 50^\circ$

}

Note that these ranges of values for velocities and dynamical timescales/ages
are in agreement with proper motion measurements e.g. from \cite{Davis2009} and
\cite{Eisloffel1994}. Each set of random projected length distributions was then
compared via a Kolmogorov-Smirnov test with the observed distribution. This
allowed us to determine the probability that the model distribution and the data
are drawn from the same parent sample. When comparing the two length
distributions based on the different distance calibration methods, we find that
they agree with a 95\,\% probability. Any models that agree worse than 10\,\%
with any of the data sets are considered bad.

 %
 %

We then investigate which models consistently lead to such a low agreement with
the observations in order to exclude parameter values for the model. The best
minimum inclination angle is about 20$^\circ$. This shows that our sample
typically does not contain many objects aligned perpendicular to the plane of
the sky. Models with a lower minimum inclination angle generate too many short
outflows, and models with a larger minimum inclination angle lack short
objects. 

The best fitting minimum/maximum velocity values are 40\,km\,s$^{-1}$ --
130\,km\,s$^{-1}$, while the age range of 4\,--20\,$\times$\,10$^3$\,yrs gives
the best agreement with the data. These values lead to a range of jet lengths
between 0.1\,pc and 2.3\,pc, in more or less the correct observed (exponentially
decreasing) distribution. We note that these dynamical lifetimes are still at
least an order of magnitude below estimates for protostellar lifetimes, which
are typically a few 10$^5$\,yrs (\cite{Hatchell2007a}).

The above  parameter values imply that our sample does not contain very young
and/or very slow moving jets. If they were very young they might have been
missed as they are still deeply embedded and thus extincted. The same applies for
the very slow moving jets, which additionally would lead to weaker shocks and
thus less bright H$_2$ emission. 

More detailed and realistic models should be tested against the available data.
In particular the speed of the jet will change over time as energy and momentum
is lost by radiation and entrainment of material. Any model should not just
reproduce the length distribution but also the distribution of 1-0\,S(1)
luminosities and the relation between jet length and brightness (see below).
However, only once the entire survey is analysed, will we have sufficient
numbers of outflows to attempt this.

\begin{figure}\centering
\includegraphics[width=8.0cm]{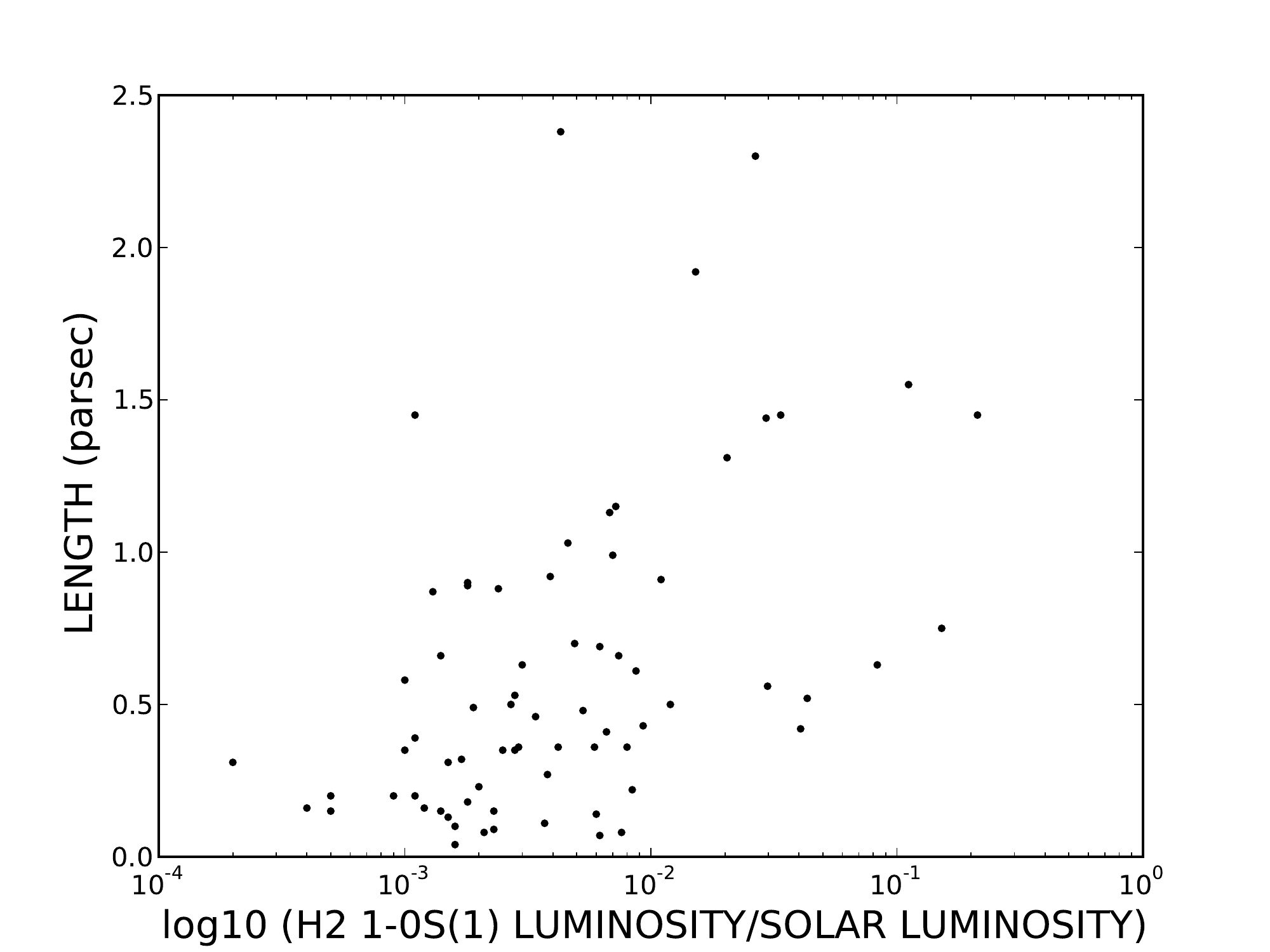}
\caption{\label{lumleng} Projected outflow length against the outflow
1-0\,S(1) luminosity based on the distance calibration method\,1. Objects
without determined distance are excluded.}   
\end{figure}
 
\subsection{Outflow length vs. luminosity}
 
In Fig.\,\ref{lumleng} we plot the 1-0\,S(1) luminosity against the projected
outflow length. The plot demonstrates that the majority of the outflows is
fainter than $10^{-2}$\,L$_\odot$ and shorter than 1\,pc in length.  However,
despite the poor statistics for bright outflows, there is a trend (corellation
coefficient $r = 0.47$) of increasing length with brightness. Essentially the
bright outflows (L$_{{\rm H}_2} > 10^{-2}$\,L$_\odot$) are on average about
twice as long as the faint (L$_{{\rm H}_2} < 10^{-2}$\,L$_\odot$) objects (1\,pc
vs. 0.5\,pc). 

Brighter integrated H$_2$ luminosities are indicative of higher surface
brightness and/or larger shock area. The former will depend on a number of
things such as shock velocity, ambient gas density, magnetic field
strength/orientation, and ionization fraction (see e.g. \cite{Khanzadyan2004}).
If the environment (clumpyness of the ISM surrounding the driving source) is the
dominating factor for the outflow luminosity, then one might expect that shorter
flows are brighter (the densest material is found closer to the star formation
side), or that there is no correlation. Hence, Fig.\,\ref{lumleng} might
indicate that brighter flows are generated by faster moving material since they
are on average longer. This is also in line with the empiric relation of source
luminosity and outflow luminosity from \cite{CarattioGaratti2008} which should
not exist if the environment plays a dominant role in determining the outflow
brightness. However, as noted above for the length distribution, more realistic
models need to be tested against the full survey data in the future.

\begin{figure}
\includegraphics[width=8.0cm]{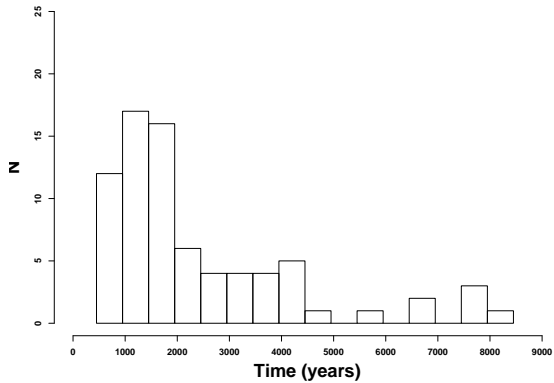}
\caption{\label{frequency} Time difference between the emission of successive
H$_2$ knots calculated based on the projected separation and an average speed of
80\,km/s for the proper motion (using distance calibration method\,1).} 
\end{figure}

\subsection{Mass ejection frequency}

In our sample there are 29 outflows which have more than one H$_2$ emission knot
on at least one side of the source. For these objects we are able to measure the
projected distances between the emission knots and thus, with an average
velocity, the time between the ejection episodes responsible for the knots. In
total we have measured 76 distances between knots in our outflows. The resulting
distributions of the ejection time differences are shown in
Fig.\,\ref{frequency}. They are based on an average speed of 80\,km\,s$^{-1}$
which is a typical speed for the jets and outflows in order to explain the
length distribution (see above).

Similarly to the projected jet length distribution, we find a larger number of
small distances between successive knots, or short times between the emission.
With increasing time/distance between knots, the number of objects
significantly decreases. There are, however, a few knots with large gaps
between them, more than expected if the general decreasing trend is to continue.

With a velocity of 80\,km\,s$^{-1}$ we find that the typical gaps between the
knots corresponds to about 10$^3$\,yrs, while the largest time gaps are about
10$^4$\,yrs. This is a variation of roughly a factor of ten, with the largest
time gaps reaching the typical dynamical jet lifetimes (see
Sect.\,\ref{sect_lengthdistr}).

Note that we only included gaps between knots which could be clearly separated
in our images (a few arcseconds -- about 10$^3$\,yrs at a typical jet speed and
a distance of 3\,kpc). On smaller scales the knot-substructure is most likely
caused or even dominated by the density structure of the ISM the jet is
interacting with and not by the ejection history itself.

We tried to model the time gap distribution in the same manner as the jet
lengths distribution (see Sect.\,\ref{sect_lengthdistr}). The same parameter
ranges for the inclination and jet velocities are used. Instead of the jet age,
we use a minimum $a_{min}$ and maximum $a_{max}$ age gap between the emission of
successive knots. 

{\centering

$1000yrs \le a_{min} \le 3000yrs$

$3000yrs \le a_{max} \le 5000yrs$

}

Such a model is able to reproduce the general trend seen in
Fig.\,\ref{frequency}, but  contrary to the jet length distribution we do not
find any model that agrees with the data above the 90\,\% level. This is most
likely due to the increased number of objects with longer time gaps, which do
not seem to follow the general trend. These large gaps can have a number of
reasons: i) One of the knots does actually not belong to the outflow; ii) The
knots are indeed emitted a long time apart; iii) We do not detect emission
in between knots due to extinction and/or they are too faint. We also need to
keep in mind that our model of constant velocity ejections is over simplistic,
and the calculated 'time gaps' between the knots are just an order of magnitude
estimate.

However, we can assume that the separation of emission knots is related to
episodes of increased mass accretion onto the central object. Numerical
simulations (e.g. from \cite{Vorobyov2006}) have shown how the mass accretion
rate can vary over time in this 'burst-mode' of star formation. These authors
obtain significant peaks in the mass accretion rate which correspond to FU-Ori
type outbursts. According to their models they occur about every
2\,$\times$\,10$^4$\,yrs. In between the mass accretion rate varies less
significant on timescales of about 1000\,yrs. Our measured time differences
hence correspond to those smaller accretion rate variations, while the total jet
lifetime (as determined from the lengths distribution above) corresponds to the
FU-Ori like eruption timescale. 

Thus, our data are in agreement with model predictions if major episodes of mass
accretion (such as FU-Ori outbursts) either trigger and/or stop the ejection of
material in a protostellar jet. Lower level accretion rate increases occur on
similar timescales as increased ejection of material in the jets (either density
changes or velocity changes will lead to new emission knots forming). However,
better statistics is required to be able to start trying to investigate how well
our data matches with different models and FU-Ori timescales. Currently ongoing
programs such as the VVV survey \cite{Minniti2010}, the YSOVAR program (e.g.
\cite{MoralesCalderon2011}) and others should soon be able to determine if the
duty cycle of FU-Ori outbursts is 10$^3$ or 10$^4$\,yrs with some certainty and
what the frequency of accretion bursts of a given strength is. 

If the larger value for the duty cycle is confirmed and FU-Ori bursts trigger a
jet ejection phase, while subsequent smaller accretion bursts are responsible
for continued emission knot formation, one would expect that statistically
outflow driving sources should show enhanced mass accretion rates compared to a
group of similar aged objects that do not drive outflows. We will test this by
investigating the driving source properties of our sample and other YSOs in the
same clouds in Paper\,III.

\section{Conclusions}\label{conclusions}


We used foreground star counts to molecular clouds associated with jets and
outflows and a comparison to the Besancon Galaxy model by \cite{Robin2003} to
determine their distances. To calibrate this method we utilised objects from the
RMS survey by \cite{Urquhart2008} which are distributed in the Galactic Plane
similar to our outflows. This method, together with the calibration allows us to
estimate distances with a typical scatter of 25\,\%. 

The majority of our detected outflows have a distance of about 3.5\,kpc,
indicating that the sight line crosses a spiral arm. The scale height of the
outflows with respect to the Galactic Plane is 30\,pc, of the same order as
massive young stars. This is in agreement with the high outflow luminosities,
and thus potential intermediate mass driving sources
(\cite{CarattioGaratti2006}) in our sample.

The outflow 1-0\,S(1) luminosities range from slightly brighter than
0.1\,L$_\odot$ to a few 10$^{-4}$\,L$_\odot$, on average an order of magnitude
brighter than in samples from nearby star forming regions. We estimate that our
sample is complete for objects brighter than 10$^{-3}$\,L$_\odot$ for distances
of up to 5\,kpc. This luminosity roughly corresponds to an HH\,211 like object
behind a K-band extinction of 1\,mag.

The luminosity distribution of the outflows shows a power law behaviour with $N
\propto L_{H_2}^{-1.9}$. With the assumption that 10\,\% of the H$_2$ flux is in
the 1-0\,S(1) line an using the empirical relation between the source bolometric
(accretion) luminosity and the outflow luminosity this translates into a
dependence of the average mass accretion rate on the final stellar mass of
$\dot{M} \propto M^{1.3 \pm 0.2}$. The total outflow luminosities also indicate
a Milky Way star formation rate (averaged over a typical jet lifetime or the
last 10$^4$\,yrs) of more than 0.4\,M$_\odot$\,yr$^{-1}$. Our sample of jets
also indicates that they are not able to provide a sizeable fraction of the
energy and momentum required to sustain the typical local levels of turbulence
in their parental clouds.

The projected jet length drops exponentially in number for longer jets, and does
not behave as a power law. The statistically corrected fraction of parsec scale
flows is 25\,\%, almost twice as high as in typical nearby star forming regions.
This is in agreement with our observed trend that more luminous outflows are
longer and the fact that the average luminosity in our sample is higher than for
outflow samples from e.g. Orion.

A simple Monte-Carlo type model of jets with speeds of 40\,--\,130\,km\,s$^{-1}$
and ages between 4\,--\,20\,$\times$\,10$^3$\,yrs can reproduce the observed
length distribution. These lifetimes are an order of magnitude below estimates
for the protostellar evolutionary phase. The model only fits the data if jets
almost perpendicular to the plane of the sky are excluded.  

Finally, we find that for typical outflow velocities the time gaps between the
ejection of larger amounts of material (resulting in groups of emission
features) are of the order of 10$^3$\,yrs. According to the burst mode of star
formation models from e.g. \cite{Vorobyov2006} the creation of the H$_2$ knots
is hence linked to low level fluctuations of the mass accretion rate and not
FU-Ori type events. Their duty cycle seems more in agreement with the total jet
lifetime, which might suggest these outburst as trigger (or stopping point or
both) of a jet ejection phase. However, better constraints of the FU-Ori duty
cycle and mechanism as well as more detailed models are required to draw any
further conclusions.

\section*{acknowledgements}

GI acknowledges a University of Kent scholarship. The United Kingdom Infrared
Telescope is operated by the Joint Astronomy Centre on behalf of the Science and
Technology Facilities Council of the U.K. The data reported here were obtained
as part of the UKIRT Service Program. 

\bibliographystyle{mn2e}
\bibliography{biblio}

\newpage
\onecolumn

\begin{appendix}

\section{MHO Properties Table}\label{appendix1}

\begin{center}

\begin{longtable}{|l|c|c|c|c|c|c|c|c|}

\caption{Summary table of the MHO properties. In cases where several MHOs belong
to the same outflow, the MHO number is labelled with an asterisk. Objects which
coincide with a RMS source are labelled with a $+$ sign. We list the MHO number,
the distance, the flux in the 1-0\,S(1) line of H$_2$, the luminosity in the
1-0\,S(1) line of H$_2$, the apparent and physical length. The numbers $1$ and
$2$ indicate values determined using our two distance calibrations. $1$ includes
the Galactic Longitude and $2$ does not. Objects where we could not determine a
distance are indicated by \dag. In these cases we use the mean distance of all
other objects to calculate luminosities and lengths.}

\label{mainresults}\\

\hline \multicolumn{1}{|c|}{\textbf{ }} & \multicolumn{1}{c|}{\textbf{Dist.}} & \multicolumn{1}{c|}{\textbf{Dist.}} & \multicolumn{1}{c|}{\textbf{F[1-0\,S(1)]}} &
\multicolumn{1}{c|}{\textbf{Lum.}}
& \multicolumn{1}{c|}{\textbf{Lum.}} & \multicolumn{1}{c|}{\textbf{Apparent}} & \multicolumn{1}{c|}{\textbf{Length}} & \multicolumn{1}{c|}{\textbf{Length}}\\

\multicolumn{1}{|c|}{\textbf{MHO}} & \multicolumn{1}{c|}{\textbf{2}} & \multicolumn{1}{c|}{\textbf{1}} & \multicolumn{1}{c|}{\textbf{ }} &
\multicolumn{1}{c|}{\textbf{2}}
& \multicolumn{1}{c|}{\textbf{1}} & \multicolumn{1}{c|}{\textbf{length}} & \multicolumn{1}{c|}{\textbf{2}} & \multicolumn{1}{c|}{\textbf{1}}\\

\multicolumn{1}{|c|}{\textbf{ }} & \multicolumn{1}{c|}{\textbf{(Kpc)}} & \multicolumn{1}{c|}{\textbf{(Kpc)}} & \multicolumn{1}{c|}{\textbf{[10E-18 W/m$^2$]}} & \multicolumn{1}{c|}{\textbf{[Solar]}}
& \multicolumn{1}{c|}{\textbf{[Solar]}} & \multicolumn{1}{c|}{\textbf{(arcsec)}} & \multicolumn{1}{c|}{\textbf{(pc)}} & \multicolumn{1}{c|}{\textbf{(pc)}} \\ \hline 
\endfirsthead

\multicolumn{9}{c}%
{{\bfseries \tablename\ \thetable{} -- continued from previous page}} \\

\hline \multicolumn{1}{|c|}{\textbf{ }} & \multicolumn{1}{c|}{\textbf{Dist.}} & \multicolumn{1}{c|}{\textbf{Dist.}} & \multicolumn{1}{c|}{\textbf{F[1-0\,S(1)]}} &
\multicolumn{1}{c|}{\textbf{Lum.}}
& \multicolumn{1}{c|}{\textbf{Lum.}} & \multicolumn{1}{c|}{\textbf{Apparent}} & \multicolumn{1}{c|}{\textbf{Length}} & \multicolumn{1}{c|}{\textbf{Length}}\\

\multicolumn{1}{|c|}{\textbf{MHO}} & \multicolumn{1}{c|}{\textbf{2}} & \multicolumn{1}{c|}{\textbf{1}} & \multicolumn{1}{c|}{\textbf{ }} &
\multicolumn{1}{c|}{\textbf{2}}
& \multicolumn{1}{c|}{\textbf{1}} & \multicolumn{1}{c|}{\textbf{length}} & \multicolumn{1}{c|}{\textbf{2}} & \multicolumn{1}{c|}{\textbf{1}}\\

\multicolumn{1}{|c|}{\textbf{ }} & \multicolumn{1}{c|}{\textbf{(Kpc)}} & \multicolumn{1}{c|}{\textbf{(Kpc)}} & \multicolumn{1}{c|}{\textbf{[10E-18 W/m$^2$]}} & \multicolumn{1}{c|}{\textbf{[Solar]}}
& \multicolumn{1}{c|}{\textbf{[Solar]}} & \multicolumn{1}{c|}{\textbf{(arcsec)}} & \multicolumn{1}{c|}{\textbf{(pc)}} & \multicolumn{1}{c|}{\textbf{(pc)}} \\ \hline

\endhead

\hline \multicolumn{9}{|r|}{{Continued on next page}} \\ \hline
\endfoot

\hline \hline
\endlastfoot

MHO 2201$^{*}$ & 3.1     & 4.1     & 405.482 & 0.1212 & 0.2125 & 73  & 1.1  & 1.45\\
MHO 2212$^{*}$ &         &         &         &        &        &     &      &     \\
MHO 2202$^{+}$ & 3.1     & 4.1     & 77.481  & 0.0231 & 0.0406 & 21  & 0.32 & 0.42\\
MHO 2203       & 2.1     & 3.1     & 278.893 & 0.0379 & 0.0832 & 42  & 0.43 & 0.63\\
MHO 2204       & 2.1     & 3.1     & 509.659 & 0.0692 & 0.1520 & 50  & 0.51 & 0.75\\
MHO 2205       & 2.1     & 3.1     & 58.429  & 0.0080 & 0.0174 & -   & -    & -   \\
MHO 2206$^{*}$ & 3.4     & 3.4     & 304.384 & 0.1100 & 0.1115 & 93  & 1.54 & 1.55\\
MHO 2207$^{*}$ &         &         &         &        &        &     &      &     \\
MHO 2208$^{*}$ &         &         &         &        &        &     &      &     \\
MHO 2209       & 3.4     & 3.4     & 25.340  & 0.0092 & 0.0093 & 26  & 0.43 & 0.43\\
MHO 2210       & 3.4     & 3.4     & 20.150  & 0.0073 & 0.0074 & 40  & 0.66 & 0.66\\
MHO 2244       & 2.9     & 3.9     & 3.005   & 0.0008 & 0.0014 & 35  & 0.50 & 0.66\\
MHO 2245       & 2.1     & 3.1     & 12.587  & 0.0017 & 0.0038 & 18  & 0.18 & 0.27\\
MHO 2246       & 2.1     & 3.1     & 7.588   & 0.0010 & 0.0023 & 10  & 0.10 & 0.15\\
MHO 2247$^{+}$ & 2.4     & 3.3     & 19.772  & 0.0035 & 0.0068 & 70  & 0.80 & 1.13\\
MHO 2248       & 2.1     & 3.1     & 3.800   & 0.0005 & 0.0011 & 26  & 0.26 & 0.39\\
MHO 2249       & 2.2     & 3.1     & 3.669   & 0.0005 & 0.0011 & 95  & 0.99 & 1.45\\
MHO 2250       & 4.3     & 4.9     & 8.288   & 0.0048 & 0.0062 & 29  & 0.61 & 0.69\\
MHO 2251$^{+}$ & 4.3     & 4.9     & 3.033   & 0.0018 & 0.0023 & 4   & 0.08 & 0.09\\
MHO 2252       & 3.0     & 4.0     & 2.406   & 0.0007 & 0.0012 & 8.5 & 0.12 & 0.16\\
MHO 2253       & 4.0     & 4.7     & 10.497  & 0.0053 & 0.0073 & -   & -    & -   \\
MHO 2254       & 3.3     & 3.9     & 9.463   & 0.0032 & 0.0046 & 54  & 0.86 & 1.03\\
MHO 2255       & 3.5     & 4.0     & 1.058   & 0.0004 & 0.0005 & 10  & 0.17 & 0.20\\
MHO 2256       & 3.5     & 4.0     & 3.435   & 0.0013 & 0.0018 & 9   & 0.15 & 0.18\\
MHO 2257       & 3.5     & 4.0     & 4.862   & 0.0018 & 0.0025 & 18  & 0.3  & 0.35\\
MHO 2258       & 3.3     & 3.9     & 4.044   & 0.0014 & 0.0019 & -   & -    & -   \\
MHO 2259       & 2.6     & 3.3     & 7.424   & 0.0015 & 0.0025 & -   & -    & -   \\
MHO 2260       & 3.7     & 4.2     & 22.339  & 0.0098 & 0.0120 & 25  & 0.45 & 0.50\\
MHO 2261       & 3.7     & 4.2     & 62.588  & 0.0273 & 0.0337 & 72  & 1.31 & 1.45\\
MHO 2262$^{+}$ & 4.7     & 4.7     & 9.502   & 0.0066 & 0.0066 & 18  & 0.41 & 0.41\\
MHO 2263       & 3.9     & 4.1     & 6.439   & 0.0031 & 0.0034 & 23  & 0.44 & 0.46\\
MHO 2264       & 3.9     & 4.1     & 15.185  & 0.0073 & 0.0080 & 18  & 0.34 & 0.36\\
MHO 2265       & 3.1     & 3.7     & 3.983   & 0.0012 & 0.0017 & 18  & 0.27 & 0.32\\
MHO 2266       & 3.4     & 3.9     & 5.976   & 0.0022 & 0.0028 & 28  & 0.46 & 0.53\\
MHO 2267       & 3.8     & 4.1     & 3.236   & 0.0015 & 0.0017 & -   & -    & -   \\
MHO 2268       & 4.4     & 4.5     & 0.108   & 0.0001 & 0.0001 & -   & -    & -   \\
MHO 2269$^{+}$ & 4.3     & 4.5     & 32.477  & 0.0191 & 0.0204 & 60  & 1.27 & 1.31\\
MHO 2270       & 3.4     & 3.8     & 1.148   & 0.0004 & 0.0005 & 8   & 0.13 & 0.15\\
MHO 2271       & 3.4     & 3.8     & 6.578   & 0.0024 & 0.0029 & 19.5& 0.33 & 0.36\\
MHO 2272       & 3.4     & 3.7     & 3.822   & 0.0014 & 0.0016 & 2.5 & 0.04 & 0.04\\
MHO 2273       & 4.0     & 4.1     & 2.535   & 0.0012 & 0.0013 & -   & -    & -   \\
MHO 2274       & 3.4     & 3.6     & 17.829  & 0.0065 & 0.0072 & 66  & 1.1  & 1.15\\
MHO 2275       & 3.4     & 3.6     & 17.181  & 0.0062 & 0.0069 & -   & -    & -   \\
MHO 2276       & 3.9     & 4.0     & 2.777   & 0.0013 & 0.0014 & 8   & 0.15 & 0.15\\
MHO 2277       & 3.6\dag & 3.7\dag & 24.144  & 0.0097 & 0.0104 & -   & -    & -   \\
MHO 2278       & 3.7\dag & 3.7\dag & 13.823  & 0.0058 & 0.0059 & 20  & 0.36 & 0.36\\
MHO 2279       & 4.3     & 4.2     & 3.023   & 0.0017 & 0.0016 & 5   & 0.1  & 0.10\\
MHO 2280$^{*}$ & 4.3     & 4.2     & 13.119  & 0.0074 & 0.0070 & 49  & 1.01 & 0.99\\
MHO 2281$^{*}$ &         &         &         &        &        &     &      &     \\
MHO 2282       & 3.5     & 3.6     & 3.561   & 0.0014 & 0.0014 & -   & -    & -   \\
MHO 2283       & 3.5     & 3.6     & 2.511   & 0.0009 & 0.0010 & 20  & 0.33 & 0.35\\
MHO 2284$^{+}$ & 3.5     & 3.6     & 3.745   & 0.0014 & 0.0015 & 18  & 0.3  & 0.31\\
MHO 2285       & 3.5     & 3.6     & 2.482   & 0.0009 & 0.0010 & 33  & 0.55 & 0.58\\
MHO 2286       & 3.4     & 3.6     & 13.093  & 0.0048 & 0.0053 & -   & -    & -   \\
MHO 2287       & 3.6     & 3.7     & 3.239   & 0.0013 & 0.0013 & -   & -    & -   \\
MHO 2288       & 3.6     & 3.7     & 10.087  & 0.0040 & 0.0042 & -   & -    & -   \\
MHO 2289       & 3.6     & 3.7     & 3.036   & 0.0012 & 0.0013 & -   & -    & -   \\
MHO 2290       & 3.6     & 3.7     & 12.735  & 0.0051 & 0.0053 & 27  & 0.47 & 0.48\\
MHO 2291       & 3.4     & 3.5     & 38.832  & 0.0144 & 0.0152 & 112 & 1.87 & 1.92\\
MHO 2292       & 5.3     & 4.9     & 7.950   & 0.0070 & 0.0060 & 6   & 0.16 & 0.14\\
MHO 2293       & 3.5     & 3.4     & 5.838   & 0.0022 & 0.0021 & -   & -    & -   \\
MHO 2294       & 3.5     & 3.4     & 1.553   & 0.0006 & 0.0006 & -   & -    & -   \\
MHO 2295       & 3.5     & 3.4     & 1.385   & 0.0005 & 0.0005 & -   & -    & -   \\
MHO 2296$^{+}$ & 3.5     & 3.4     & 2.170   & 0.0008 & 0.0008 & -   & -    & -   \\
MHO 2297       & 3.3     & 3.2     & 1.326   & 0.0004 & 0.0004 & 10  & 0.16 & 0.16\\
MHO 2298       & 4.1\dag & 3.7\dag & 30.286  & 0.0156 & 0.0130 & -   & -    & -   \\
MHO 2299       & 3.0\dag & 3.7\dag & 4.781   & 0.0014 & 0.0021 & -   & -    & -   \\
MHO 2436       & 4.1\dag & 3.7\dag & 10.940  & 0.0057 & 0.0047 & -   & -    & -   \\
MHO 2437       & 4.5     & 4.1     & 28.685  & 0.0181 & 0.0150 & -   & -    & -   \\
MHO 2438       & 4.8     & 4.3     & 5.005   & 0.0036 & 0.0028 & -   & -    & -   \\
MHO 2439       & 4.1     & 3.7     & 5.294   & 0.0027 & 0.0022 & -   & -    & -   \\
MHO 2440       & 4.1     & 3.7     & 6.636   & 0.0034 & 0.0028 & 20  & 0.39 & 0.35\\
MHO 2441       & 4.1\dag & 3.7\dag & 19.608  & 0.0104 & 0.0084 & 12  & 0.24 & 0.22\\
MHO 2442       & 4.2     & 3.8     & 13.718  & 0.0076 & 0.0062 & -   & -    & -   \\
MHO 2443       & 4.9     & 4.3     & 7.008   & 0.0052 & 0.0040 & -   & -    & -   \\
MHO 2444       & 4.8     & 4.2     & 11.168  & 0.0080 & 0.0062 & 3.5 & 0.08 & 0.07\\
MHO 2445       & 4.8     & 4.3     & 3.482   & 0.0025 & 0.0020 & 11  & 0.26 & 0.23\\
MHO 2446       & 3.8     & 3.4     & 3.608   & 0.0016 & 0.0013 & -   & -    & -   \\
MHO 2447       & 3.8     & 3.4     & 6.586   & 0.0030 & 0.0024 & -   & -    & -   \\
MHO 2448       & 4.9     & 4.3     & 7.518   & 0.0056 & 0.0043 & 115 & 2.72 & 2.38\\
MHO 2449       & 4.9     & 4.3     & 3.770   & 0.0028 & 0.0021 & 4   & 0.09 & 0.08\\
MHO 2450       & 4.9     & 4.3     & 0.289   & 0.0002 & 0.0002 & 15  & 0.35 & 0.31\\
MHO 2451       & 4.9     & 4.3     & 8.806   & 0.0065 & 0.0050 & -   & -    & -   \\
MHO 2452       & 4.9     & 4.3     & 0.719   & 0.0005 & 0.0004 & -   & -    & -   \\
MHO 2453       & 4.9     & 4.3     & 3.087   & 0.0023 & 0.0018 & 43  & 1.01 & 0.89\\
MHO 2454       & 5.4     & 4.6     & 45.166  & 0.0403 & 0.0298 & 25  & 0.65 & 0.56\\
MHO 2455       & 5.4     & 4.6     & 5.353   & 0.0048 & 0.0035 & -   & -    & -   \\
MHO 2456       & 3.6\dag & 3.7\dag & 8.493   & 0.0035 & 0.0037 & -   & -    & -   \\
MHO 3200       & 3.0\dag & 3.7\dag & 68.260  & 0.0195 & 0.0294 & 80  & 1.18 & 1.44\\
MHO 3201       & 3.0\dag & 3.7\dag & 3.534   & 0.0010 & 0.0015 & 7   & 0.1  & 0.13\\
MHO 3202       & 3.6     & 3.5     & 6.238   & 0.0026 & 0.0024 & 51  & 0.9  & 0.88\\
MHO 3203       & 3.7     & 3.6     & 3.383   & 0.0015 & 0.0014 & -   & -    & -   \\
MHO 3204       & 3.7     & 3.6     & 3.304   & 0.0014 & 0.0013 & 50  & 0.91 & 0.87\\
MHO 3205       & 3.6     & 3.5     & 2.275   & 0.0009 & 0.0009 & 12  & 0.21 & 0.20\\
MHO 3206       & 3.3     & 3.3     & 3.439   & 0.0012 & 0.0011 & -   & -    & -   \\
MHO 3207       & 3.3     & 3.3     & 4.159   & 0.0014 & 0.0014 & -   & -    & -   \\
MHO 3208       & 3.3     & 3.3     & 1.440   & 0.0005 & 0.0005 & -   & -    & -   \\
MHO 3209       & 3.3     & 3.3     & 5.022   & 0.0017 & 0.0017 & -   & -    & -   \\
MHO 3210       & 3.6     & 3.5     & 2.853   & 0.0012 & 0.0011 & 12  & 0.21 & 0.20\\
MHO 3211       & 3.6     & 3.5     & 29.166  & 0.0117 & 0.0110 & 54  & 0.94 & 0.91\\
MHO 3212       & 3.6     & 3.5     & 2.527   & 0.0010 & 0.0010 & -   & -    & -   \\
MHO 3213       & 3.0     & 3.0     & 6.857   & 0.0019 & 0.0019 & 34  & 0.49 & 0.49\\
MHO 3214       & 3.9\dag & 3.7\dag & 6.214   & 0.0029 & 0.0027 & 28  & 0.53 & 0.50\\
MHO 3215       & 3.9\dag & 3.7\dag & 4.387   & 0.0021 & 0.0019 & -   & -    & -   \\
MHO 3216       & 3.1     & 4.1     & 82.356  & 0.0245 & 0.0432 & 26  & 0.39 & 0.52\\
MHO 3217       & 2.2     & 3.2     & 15.238  & 0.0022 & 0.0049 & 45  & 0.47 & 0.70\\
MHO 3218       & 2.2     & 3.2     & 5.549   & 0.0008 & 0.0018 & 58  & 0.61 & 0.90\\
MHO 3219       & 2.2     & 3.2     & 12.049  & 0.0017 & 0.0039 & 59  & 0.62 & 0.92\\
MHO 3220       & 2.2     & 3.2     & 8.344   & 0.0012 & 0.0027 & -   & -    & -   \\
MHO 3221       & 3.2     & 4.1     & 16.329  & 0.0053 & 0.0086 & -   & -    & -   \\
MHO 3222       & 2.5     & 3.4     & 10.380  & 0.0020 & 0.0037 & 6.5 & 0.08 & 0.11\\
MHO 3223       & 1.9     & 2.9     & 3.129   & 0.0004 & 0.0008 & -   & -    & -   \\
MHO 3224       & 2.6     & 3.5     & 1.700   & 0.0004 & 0.0006 & -   & -    & -   \\
MHO 3225       & 2.6     & 3.5     & 0.964   & 0.0002 & 0.0004 & -   & -    & -   \\
MHO 3226       & 2.6     & 3.5     & 3.466   & 0.0007 & 0.0013 & -   & -    & -   \\
MHO 3227       & 2.6     & 3.5     & 0.586   & 0.0001 & 0.0002 & -   & -    & -   \\
MHO 3228       & 2.0     & 3.1     & 2.092   & 0.0003 & 0.0006 & -   & -    & -   \\
MHO 3229       & 4.1     & 4.2     & 3.135   & 0.0017 & 0.0017 & -   & -    & -   \\
MHO 3230       & 2.0     & 3.1     & 5.628   & 0.0007 & 0.0017 & -   & -    & -   \\
MHO 3231       & 2.0     & 3.1     & 2.024   & 0.0002 & 0.0006 & -   & -    & -   \\
MHO 3232       & 2.0     & 3.1     & 1.201   & 0.0001 & 0.0004 & -   & -    & -   \\
MHO 3233       & 2.2     & 3.3     & 1.428   & 0.0002 & 0.0005 & -   & -    & -   \\
MHO 3234       & 3.1\dag & 3.7\dag & 3.940   & 0.0012 & 0.0017 & -   & -    & -   \\
MHO 3235       & 3.7     & 3.9     & 8.045   & 0.0034 & 0.0039 & -   & -    & -   \\
MHO 3236       & 2.9     & 3.3     & 0.913   & 0.0002 & 0.0003 & -   & -    & -   \\
MHO 3237       & 4.1     & 4.2     & 7.729   & 0.0041 & 0.0042 & 18  & 0.36 & 0.36\\
MHO 3238       & 3.2     & 3.4     & 8.895   & 0.0029 & 0.0033 & -   & -    & -   \\
MHO 3239       & 3.2     & 3.4     & 15.135  & 0.0049 & 0.0055 & -   & -    & -   \\
MHO 3240       & 3.6     & 3.8     & 60.245  & 0.0249 & 0.0266 & 126 & 2.22 & 2.30\\
MHO 3241       & 3.1     & 3.1     & 28.141  & 0.0083 & 0.0087 & 40  & 0.6  & 0.61\\
MHO 3242$^{+}$ & 3.4     & 3.4     & 16.002  & 0.0058 & 0.0059 & -   & -    & -   \\
MHO 3243       & 3.4     & 3.4     & 8.172   & 0.0030 & 0.0030 & 38  & 0.63 & 0.63\\
MHO 3244       & 3.4     & 3.4     & 8.433   & 0.0030 & 0.0031 & -   & -    & -   \\
MHO 3246       & 3.1\dag & 3.7\dag & 17.575  & 0.0053 & 0.0076 & 4.5 & 0.07 & 0.08\\
     
\end{longtable}
\end{center}

\clearpage
\newpage
\begin{landscape}

\section{Corrected MHO Properties Table}\label{appendix2}

\begin{center}
\begin{longtable}{|l|l|l|c|c|c|c|l|l|l|}

\caption{Properties of the five MHOs for which we identified a different driving
source candidate compared to Paper\,I. An asterisk indicates MHOs which belong
to the same outflow.}

\label{outflow_table}\\

\hline \multicolumn{1}{|c|}{\textbf{MHO}} & \multicolumn{1}{c|}{\textbf{RA}} & \multicolumn{1}{c|}{\textbf{DEC}} & \multicolumn{1}{c|}{\textbf{F[1-0\,S(1)]}} &
\multicolumn{1}{c|}{\textbf{Flux error}}
& \multicolumn{1}{c|}{\textbf{length}} & \multicolumn{1}{c|}{\textbf{position angle}} & \multicolumn{1}{c|}{\textbf{possible}} & \multicolumn{1}{c|}{\textbf{source RA}} & \multicolumn{1}{c|}{\textbf{source DEC}}\\
\multicolumn{1}{|c|}{\textbf{ }} & \multicolumn{1}{c|}{\textbf{(J2000)}} & \multicolumn{1}{c|}{\textbf{(J2000)}} & \multicolumn{1}{c|}{\textbf{[10E-18 W/m$^2$]}} & \multicolumn{1}{c|}{\textbf{[10E-18 W/m$^2$]}}
& \multicolumn{1}{c|}{\textbf{(arcsec)}} & \multicolumn{1}{c|}{\textbf{(degrees)}} & \multicolumn{1}{c|}{\textbf{source}} & \multicolumn{1}{c|}{\textbf{(J2000)}} & \multicolumn{1}{c|}{\textbf{(J2000)}}\\ \hline 
\endfirsthead

\multicolumn{10}{c}%
{{\bfseries \tablename\ \thetable{} -- continued from previous page}} \\
\hline \multicolumn{1}{|c|}{\textbf{MHO}} & \multicolumn{1}{c|}{\textbf{RA}} & \multicolumn{1}{c|}{\textbf{DEC}} & \multicolumn{1}{c|}{\textbf{F[1-0\,S(1)]}} &
\multicolumn{1}{c|}{\textbf{Flux error}}
& \multicolumn{1}{c|}{\textbf{length}} & \multicolumn{1}{c|}{\textbf{position angle}} & \multicolumn{1}{c|}{\textbf{possible}} & \multicolumn{1}{c|}{\textbf{source RA}} & \multicolumn{1}{c|}{\textbf{source DEC}}\\
\multicolumn{1}{|c|}{\textbf{ }} & \multicolumn{1}{c|}{\textbf{(J2000)}} & \multicolumn{1}{c|}{\textbf{(J2000)}} & \multicolumn{1}{c|}{\textbf{[10E-18 W/m$^2$]}} & \multicolumn{1}{c|}{\textbf{[10E-18 W/m$^2$]}}
& \multicolumn{1}{c|}{\textbf{(arcsec)}} & \multicolumn{1}{c|}{\textbf{(degrees)}} & \multicolumn{1}{c|}{\textbf{source}} & \multicolumn{1}{c|}{\textbf{(J2000)}} & \multicolumn{1}{c|}{\textbf{(J2000)}}\\ \hline 
\endhead

\hline \multicolumn{10}{|r|}{{Continued on next page}} \\ \hline
\endfoot

\hline \hline
\endlastfoot

MHO 2271       & 18:35:31.7 & -08:52:17 & 6.578   & 0.465   & 19.5 & 102 & G023.2293-00.5289 & 18:35:30.4 & -08:52:13 \\
MHO 2272       & 18:35:51.3 & -08:41:13 & 3.822   & 0.239   & 2.5  & 135 & G023.4319-00.5212 & 18:35:51.4 & -08:41:10 \\
MHO 2276       & 18:35:22.9 & -07:19:17 & 2.777   & 0.165   & 8    & 180 & G024.5919+00.2119 & 18:35:22.9 & -07:19:10 \\
MHO 2280$^{*}$ & 18:38:55.4 & -06:52:37 & 6.770   & 1.073   & 49   & 145 & G025.3846-00.3724 & 18:38:56.5 & -06:53:02 \\
MHO 2281$^{*}$ & 18:38:57.3 & -06:53:15 & 6.350   & 0.620   & *    & *   & *                 & *          & *         \\

\end{longtable}
\end{center}
\end{landscape}

\clearpage
\newpage

\section{Corrected MHO images}\label{appendix3}

\begin{center}
\begin{longtable}{|r|c|p{3.5in}|}

\caption{Finding charts of the MHOs where we identified different driving
sources compared to Paper\,I. An asterisk indicates MHOs which belong to the
same outflow.}

\label{images_table} \\

\hline \multicolumn{1}{|c|}{\textbf{MHO}} & \multicolumn{1}{c|}{\textbf{Image}} & \multicolumn{1}{c|}{\textbf{Comments}} \\ \hline 
\endfirsthead

\multicolumn{3}{c}%
{{\bfseries \tablename\ \thetable{} -- continued from previous page}} \\
\hline \multicolumn{1}{|c|}{\textbf{MHO}} & \multicolumn{1}{c|}{\textbf{Image}} & \multicolumn{1}{c|}{\textbf{Comments}} \\ \hline 
\endhead

\hline \multicolumn{3}{|r|}{{Continued on next page}} \\ \hline
\endfoot

\hline \hline
\endlastfoot

MHO 2271       & \parbox[c]{2in}{\includegraphics[width=2in]{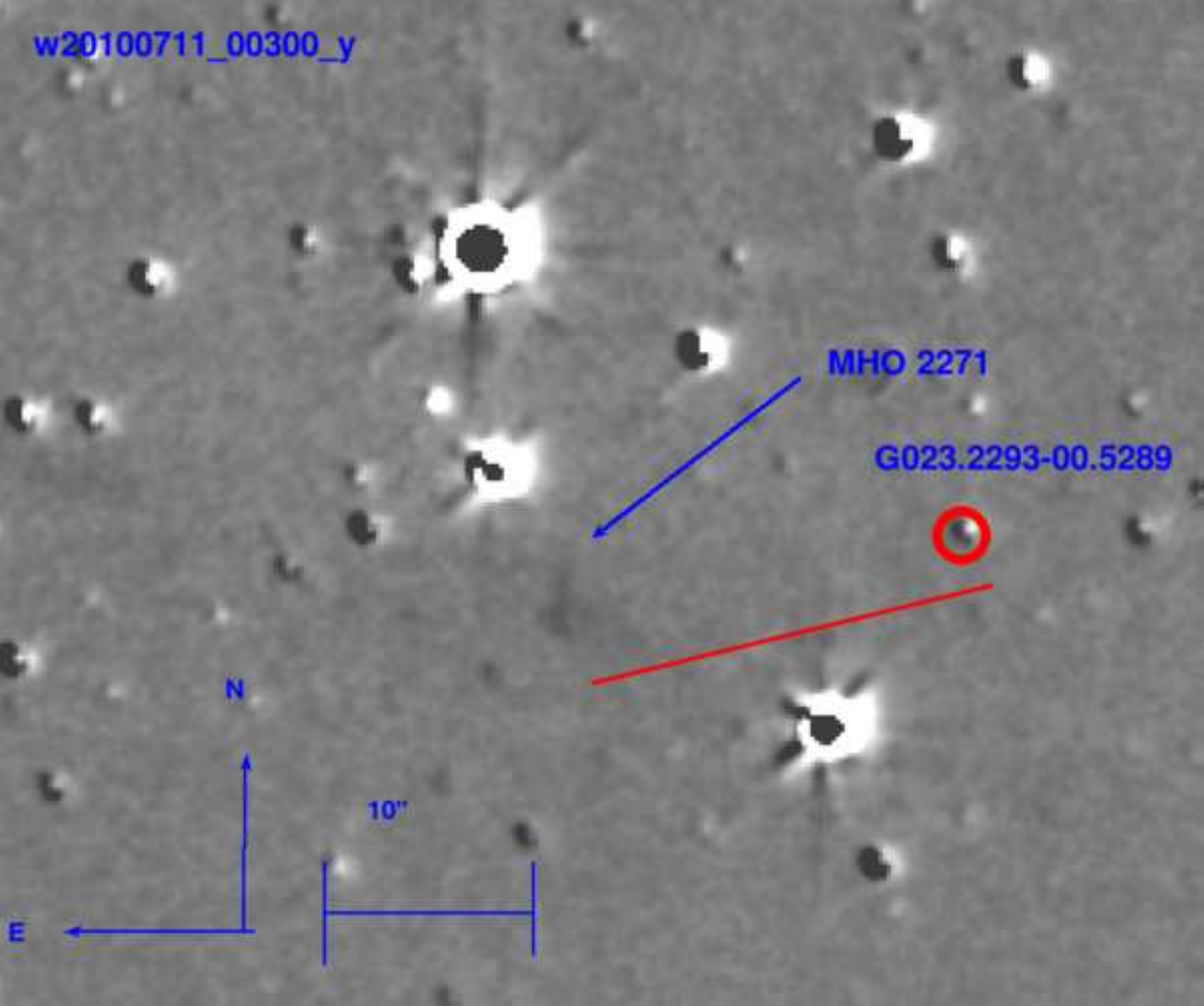}} & A faint bow shock like emission to the South East of candidate source Glimpse G023.2293-00.5289.\\ \hline
MHO 2272       & \parbox[c]{2in}{\includegraphics[width=2in]{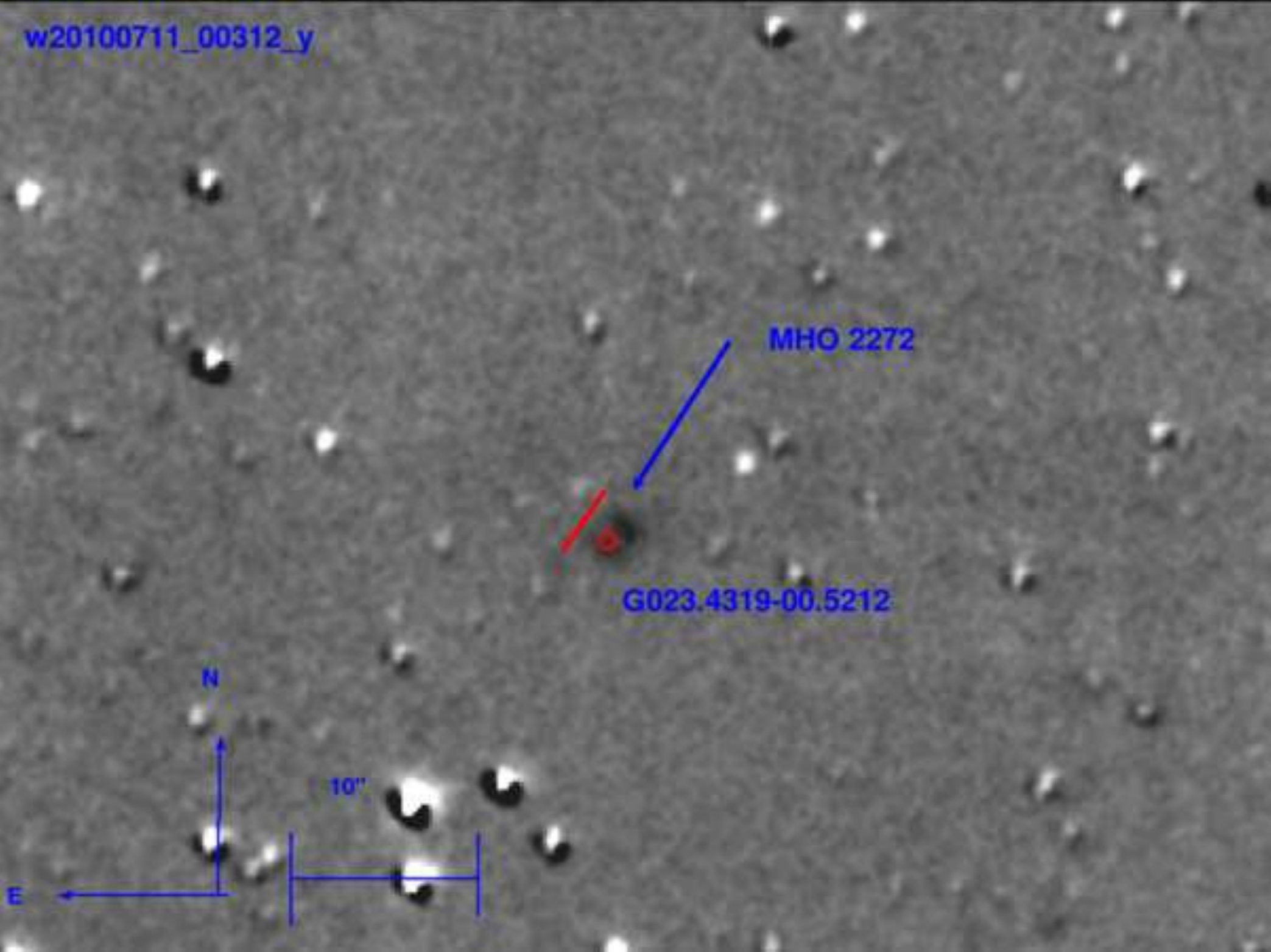}} & Two compact knots with candidate source Glimpse G023.4319-00.5212 in the middle.\\ \hline
MHO 2276       & \parbox[c]{2in}{\includegraphics[width=2in]{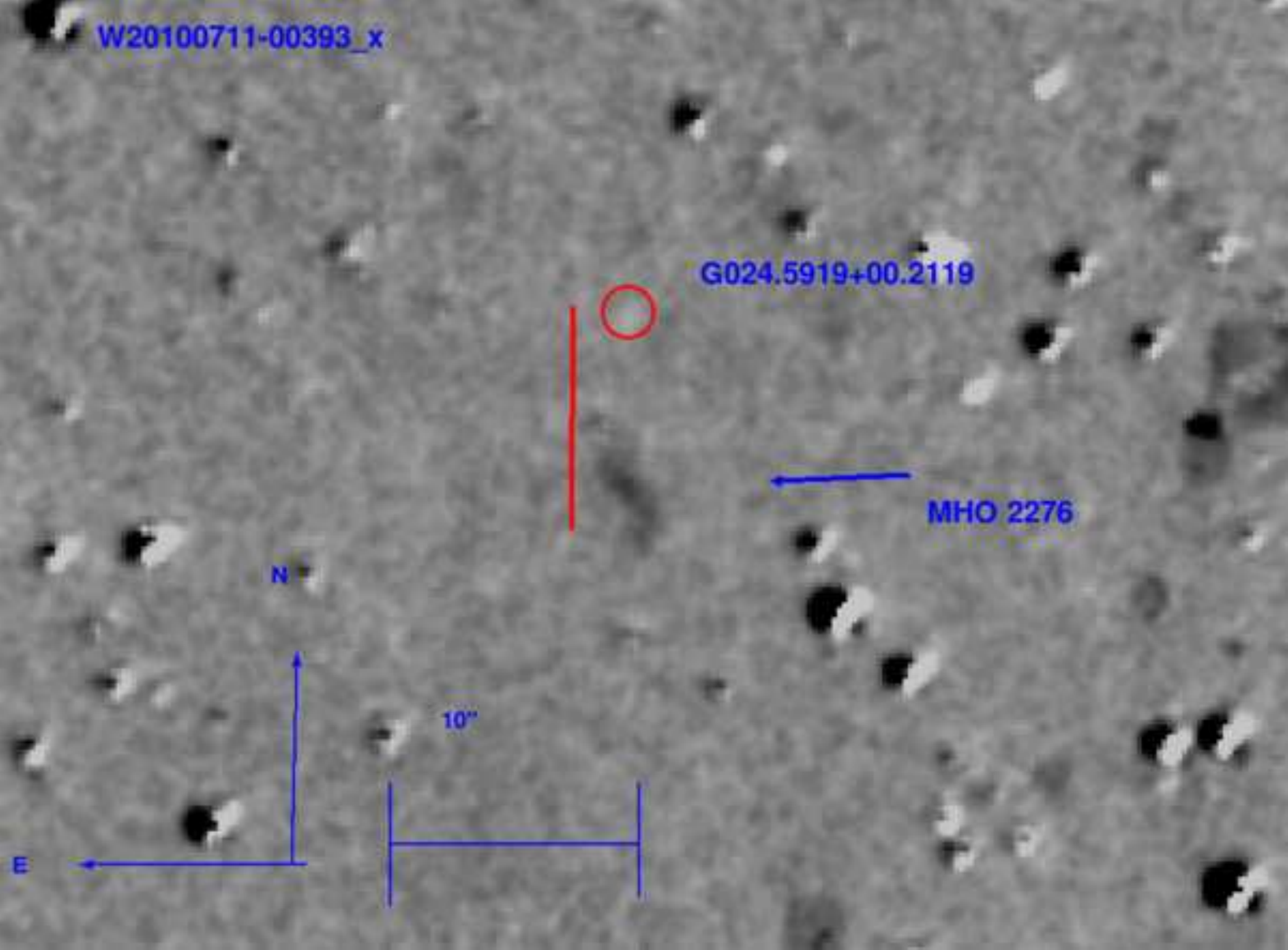}} & A faint elongated emission knot South of candidate source Glimpse G024.5919+00.2119.\\ \hline
MHO 2280$^{*}$ & \parbox[c]{2in}{\includegraphics[width=2in]{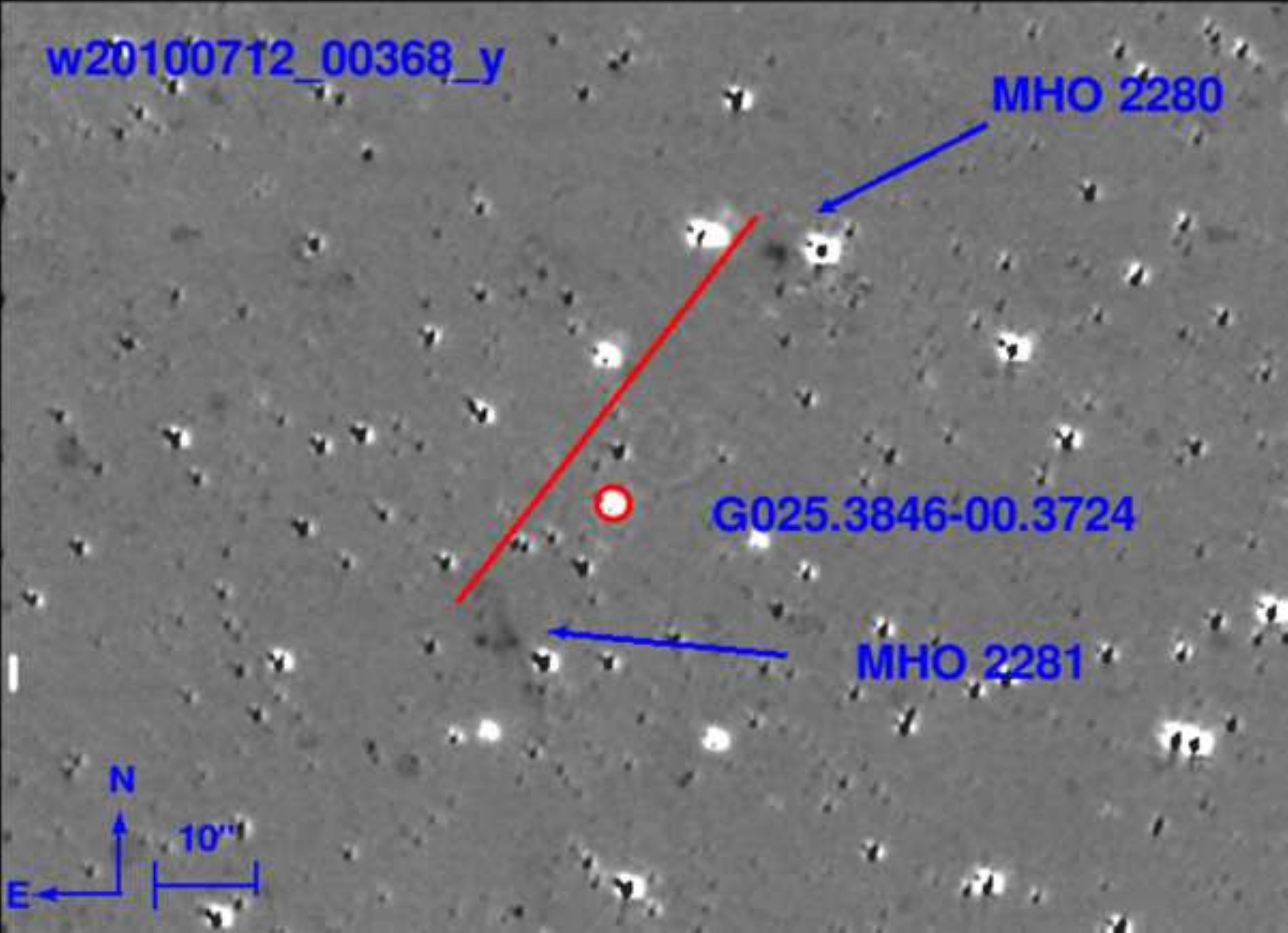}} & A bright emission knot aligned with MHO\,2281 and most likely driven by candidate source Glimpse G025.3846-00.3724. \\ \hline
MHO 2281$^{*}$ & \parbox[c]{2in}{\includegraphics[width=2in]{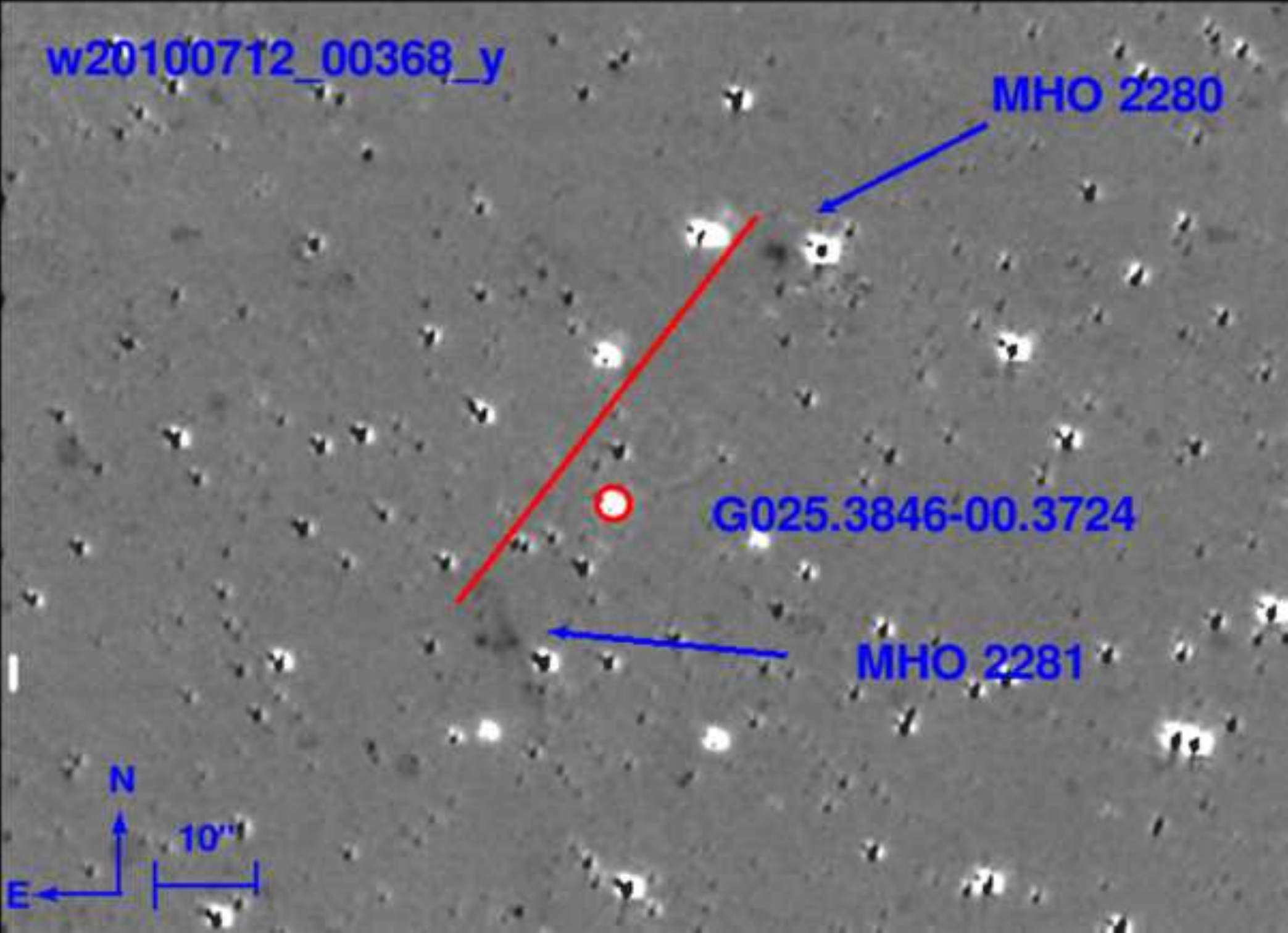}} & Extended bright, partly diffuse emission knot that is part of the same flow as MHO\,2280 that is driven by candidate source Glimpse G025.3846-00.3724.\\ \hline

\end{longtable}
\end{center}

\clearpage
\newpage

\section{Luminosity functions}\label{appendix4}

\begin{figure*}
\includegraphics[width=8.0cm]{lum2.pdf} \hfill
\includegraphics[width=8.0cm]{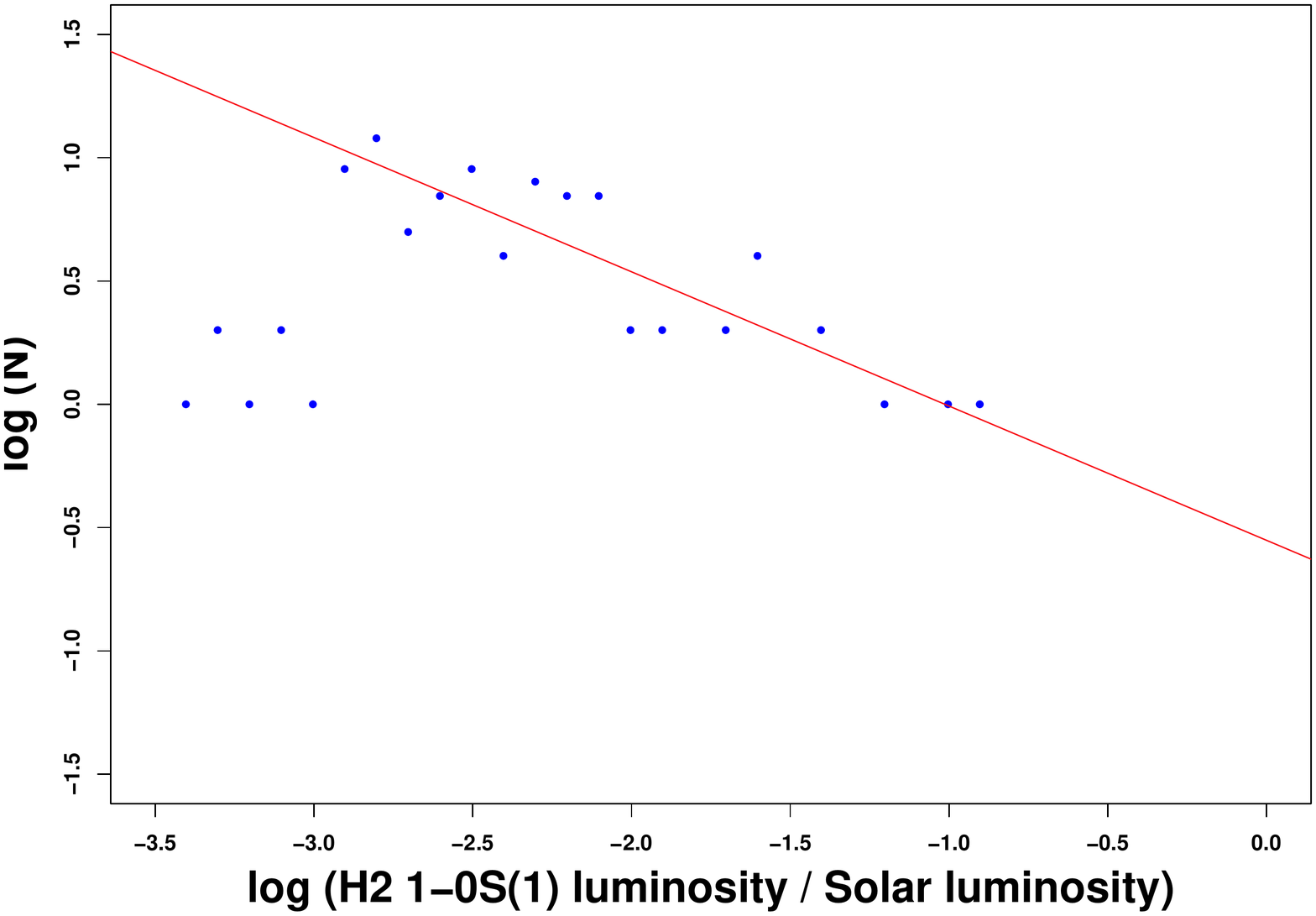} \\
\includegraphics[width=8.0cm]{Rand_Distr_Lum_2.pdf} \hfill
\includegraphics[width=8.0cm]{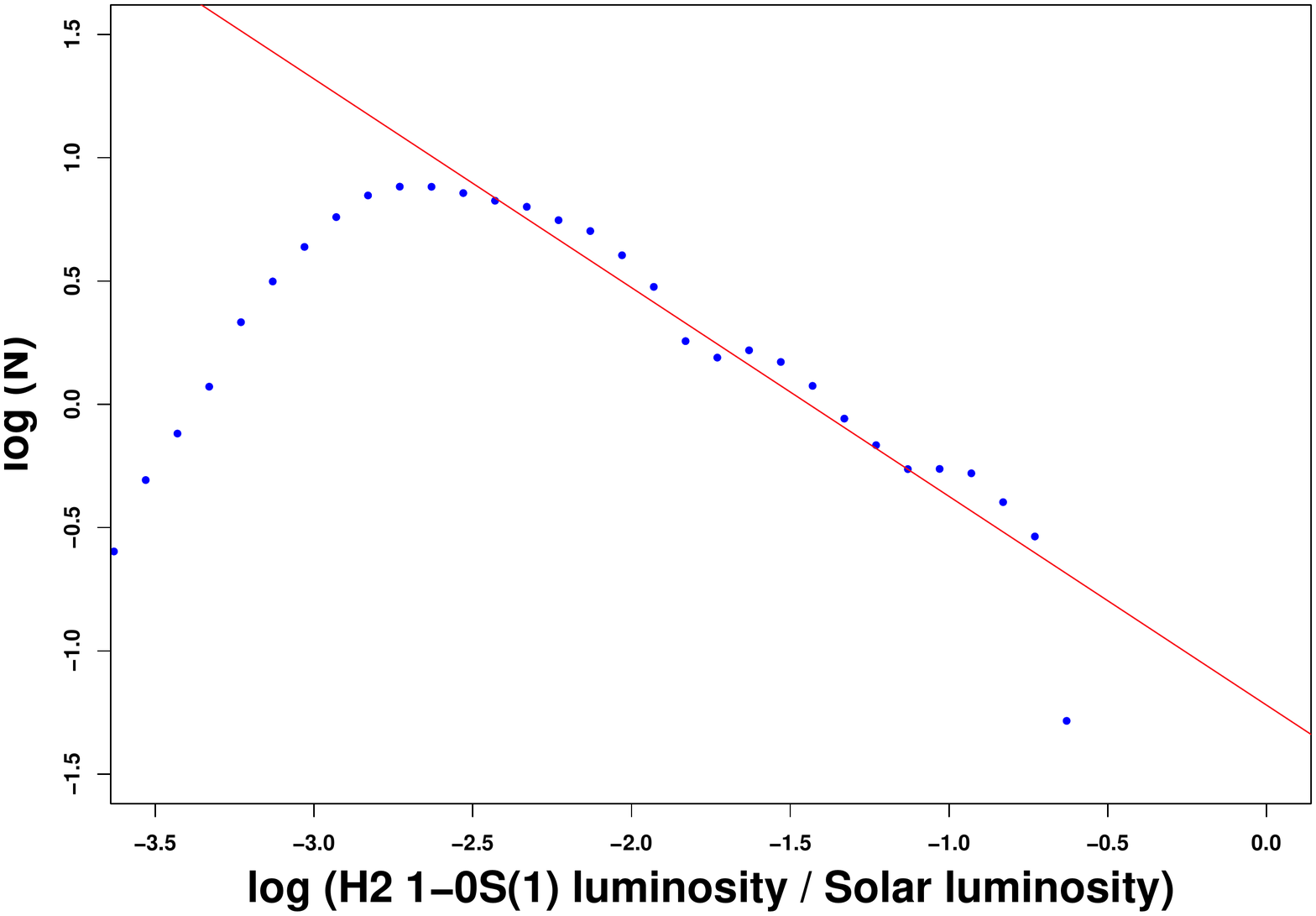} \\
\includegraphics[width=8.0cm]{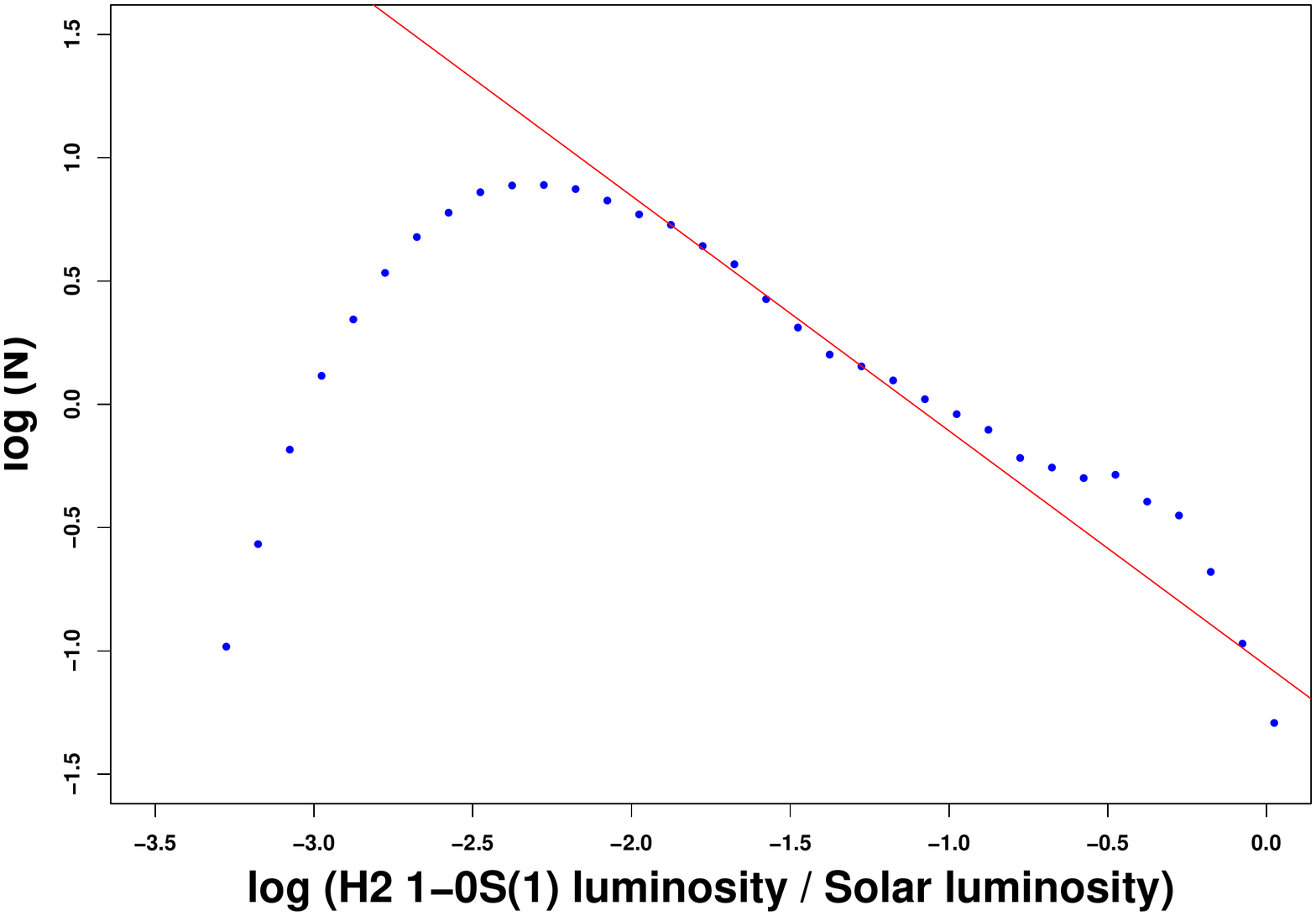} \hfill
\includegraphics[width=8.0cm]{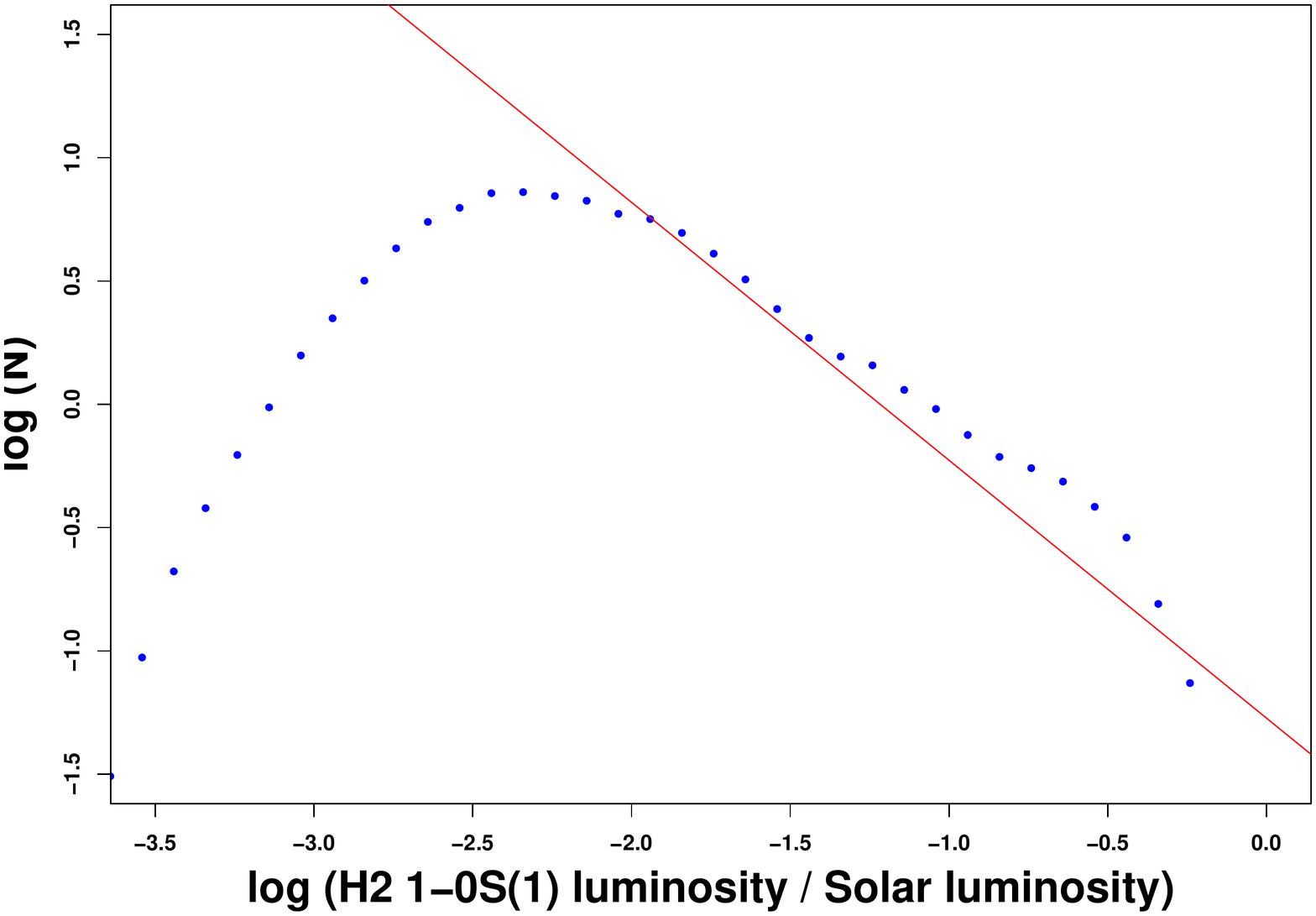}

\caption{\label{lum_appendix} 1-0\,S(1) Luminosity functions of our outflows.
{\bf Top:} Actual data; {\bf Middle:} Statistically corrected for distance
uncertainties; {\bf Bottom:} Statistically corrected for distance uncertainties
and extinction. The left column uses distance calibration method\,1, the right
column method\,2. For the statistical correction each outflow has been placed
$N=10000$ times into  the histogram (see text for details). All objects with no
measured distance are excluded, as are objects below the flux completeness
limit. The slopes for the original data are -0.5 to -0.7, depending on the bin
size. After the statistical correction the slopes are indistinguishable and have
a value of -0.9. } \end{figure*}

\end{appendix}

\label{lastpage}

\end{document}